\documentclass[12pt]{article}

\usepackage{enumerate}
\usepackage{natbib}
\usepackage{url} 

\newcommand{\blind}{1}

\usepackage{amsfonts,amssymb,amsthm}
\usepackage{geometry}
\usepackage{booktabs}  
\usepackage{xcolor}
\usepackage{natbib}
\usepackage{subfigure}
\usepackage{blkarray}
\usepackage{float}
\newcommand{\pkg}[1]{{\fontseries{b}\selectfont #1}}
\let\proglang=\textsf

\usepackage{amsmath}
\usepackage{amsfonts}
\usepackage{bm}
\usepackage{amssymb, color}
\usepackage{graphicx}
\usepackage{fullpage}
\usepackage{natbib}
\usepackage{amsthm}
\usepackage{dsfont}
\usepackage{multirow}

\newcounter{lm}

\newtheorem{lemma}[lm]{Lemma}

\DeclareMathOperator*{\argmax}{arg\,max}
\DeclareMathOperator*{\argmin}{arg\,min}

\usepackage{xspace}

\newcommand{\cov}{COVID-19\xspace}
\newcommand{\approach}{\texttt{CRFTIW}\xspace}
\newcommand{\TI}{TI\xspace}
\newcommand{\tit}{Translation-invariant functional clustering on \cov deaths adjusted on population risk factors}

\usepackage{authblk}
\begin{document}

\def\spacingset#1{\renewcommand{\baselinestretch}%
{#1}\small\normalsize} \spacingset{1}


\if1\blind
{
  \title{\bf \tit}
  \author[1]{Amay Cheam}
\author[2]{Marc Fredette}
\author[3]{Matthieu Marbac}
\author[4]{Fabien Navarro}
\affil[1,2]{HEC Montreal, Quebec, Canada}
\affil[3]{Univ. Rennes, Ensai, CNRS, CREST - UMR 9194, F-35000 Rennes, France}
\affil[4]{SAMM, Universit\'e Paris 1 Panth\'eon-Sorbonne, Paris, France}
  \maketitle
} \fi

\if0\blind
{
  \bigskip
  \bigskip
  \bigskip
  \begin{center}
    {\LARGE\bf \tit}
\end{center}
  \medskip
} \fi

\bigskip
\begin{abstract}

The \cov pandemic has taken the world by storm with its high infection rate. Investigating its geographical disparities has paramount interest in order to gauge its relationships with political decisions, economic indicators, or mental health. This paper focuses on clustering the daily death rates reported in several regions of Europe and the United States over seventeen months. Several methods have been developed to cluster such functional data. However, these methods are not translation-invariant and thus cannot handle different times of arrivals of the disease, nor can they consider external covariates and so are unable to adjust for the population risk factors of each region. 
We propose a novel three steps clustering method to circumvent these issues. As a first step, feature extraction is performed by translation-invariant wavelet decomposition which permits to deal with the different onsets. As a second step, single-index regression is used to neutralize disparities caused by population risk factors. As a third step, a nonparametric mixture is fitted on the regression residuals to achieve the region clustering. 
\end{abstract}

\noindent%
{\it Keywords:}  Functional data; Mixture models; Semiparametric models; Single-index regression; Wavelets;
\vfill

\section{Introduction}
In March of 2020, the World Health Organization (WHO) declared pandemic status for the novel coronavirus SARS-Cov-2, denoted \cov, indicating that it has reached a critical level of spread and severity worldwide. 
The global nature of \cov pandemic has resulted in plenty of heterogeneity of the data, aggravated further by lack of prior knowledge or coordinated mitigation strategies which impeded research efforts. For instance, the assumption that the first occurrence emerged concurrently everywhere is improper. Additionally, the number of confirmed cases depends on the amount of tests that are being performed in a region. Hence, a region that has tested very few people can only report very few confirmed cases. Alternatively, the number of \cov deaths is more systematically recorded: countries are asked to follow the `cause of death' classifications from the WHO's International Classification of Diseases guidelines \citep{WHOdisease}.
Though each country is responsible to provide their own guidance on how and when \cov deaths should be recorded, this metric remains more reliable. 
Undoubtedly, the rapid propagation of this acute infectious respiratory disease has posed governmental challenges.
Government responses to contain the virus's spread were multiple (social distancing, travel restrictions, lockdowns, etc.) and their efficiency needs to be investigated.
To better understand this virus, it is profoundly useful to cluster regions similarly affected by \cov.

This paper focuses on the investigation of spatial disparities of \cov by considering the daily \cov deaths recorded over seventeen months. 
Previous works consider this aim for analyzing a specific region \citep{cioban2021spatial} and propose to perform clustering by using geometrical methods (see \citet{bullock2020mapping} and the reference inside). 
In this context, \citet{bucci2021clustering} propose to use a Bayesian nonparametric approach with a Gaussian mixture model to cluster regions in Europe, while \citet{tang2020functional} propose to use classical functional cluster analysis to cluster the continental states of the United States of America  (USA). Alternatively, \citet{chen2020clustering} propose to use a non-negative matrix factorization (NMF) followed by a $K$-means clustering procedure applied on the coefficients of the NMF basis to cluster the continental states of the USA based on the new daily confirmed case counts. These approaches perform well when the study area is focused on a specific region. However, in this paper, we focus on clustering regions of the European Union (EU) and the USA. When considering regions of Europe and North America, a difficulty arises: \cov outbreaks started at different times. The misalignment of the first occurrence between regions should not be neglected, whether between continents or within a country. 
To circumvent this issue, geographical disparities of the \cov impact can be investigated by geometric clustering approaches based on dissimilarity matrices that are not sensitive to time shifting. In this context, the dynamic time warping (DTW) approach is the most popular method to obtain such dissimilarity matrix. Then, hierarchical clustering \citep{park2020clustering} or spectral clustering can be performed by considering this matrix \citep{allem2020spectral}.  
Note that \citet{michael2021learning}  the time shifting between the curves in a semi-supervised approach to learn different curve profile in UE and use these patterns to predict the evolution of the disease in the USA. 
 Another problem to acknowledge is that the mortality occurs at different rates under different population risk factors \citep{Williamson2020}. Hence the necessity to adjust these region-specific risk factors is intrinsic to allow regions to be compared fairly. Indeed, if these factors are neglected during clustering, then one observed a correlation between the estimated clusters and the risk factors \citep{ramirez2021spatial}. Furthermore, by adjusting the population risk factors, we are able to detect regions more susceptible to \cov and perhaps identify the disparity factors between clusters. For instance, it allows for a retrospective assessment of the effectiveness and the quality of government responses, a concurrent analysis of the economic indicators, and a prospective perception of mental health during this unprecedented period.


Traditionally, clustering may be achieved through finite mixture models \citep{McLachlan:04}. When the family of distributions for each cluster is unknown, nonparametric mixtures can be considered to avoid unjustified parametric assumptions \citep{chauveauSurveys2015}. A classical approach among these methods is to define the density of each mixture component as a product of univariate densities \citep{hallAOS2003,kasahara2014non}. 
The data we analyze are functional and thus raise the problem of data dimension \citep{ferraty06,ramsay2007applied}. To circumvent the curse of dimensionality, many model-based clustering approaches approximate the observed functions in some functional basis then perform clustering on the coefficients related to the basis (see the review of \citet{jacques14} or \citet{cheam2020}). For instance,  \cite{bouveyron2015discriminative} proposed approximating the curves into a Fourier basis expansion coefficients, then perform clustering on the obtained coefficients with a Gaussian mixture. Alternatively, feature extraction could be accomplished via an orthogonal wavelet basis \citep{antoniadis2013clustering}.  A further issue raised by this type of data is that of curve alignment. This has been addressed by previous works that do not tackle clustering \citep{kneip1992statistical,wang1997alignment,ramsay1998curve}. Recently, this issue has been considered for clustering in  distance-based \citep{paparrizos2015k} and model-based \citep{chudova2003translation,gaffney2005joint,liu09} frameworks.

In this article, we propose a novel three-step approach that circumvents the issues of clustering regions with respect to the \cov dataset: the varying times of arrivals of the virus and the need to incorporate the population risk factors. 
This approach is named \emph{Clustering Regression residuals of Features given by Translation Invariant Wavelets} (\approach). The first step of \approach consists of feature extraction using a multiscale approach based on translation-invariant (\TI) wavelets \citep{Coifman1995}, which allows  the shifted onsets of \cov to be tackled by avoiding any pre-processing step for curve alignment (see \cite{wang1997alignment} and the references cited in \citet[Section 2.3]{jacques14}). The objective to construct clusters that are invariant to time-shifts is somewhat different from conventional clustering in that it permits us to answer slightly different scientific questions about the data. Standard clustering (no time-shifts) will identify regions that peak at the same period, while \TI clustering recovers regions that behave in similar patterns that unravel across time.
The features are defined as the logarithm of the norm of the \TI wavelet coefficients at each scale. The second step of \approach integrates the population characteristics with a single-index regression of the features on the population risk factors. This approach has the benefits of the nonparametric regression but does not suffer from the curse of dimensionality. We show  that the residuals of the regression preserve the cluster information. 
As the third step of \approach,  clustering of the regions is achieved by fitting a nonparametric mixture on the regression residuals. The only assumption made at this step is to define the density of each component as a product of univariates densities. 
The proposed approach has differences with the approach of \citet{gaffney2005joint} despite both methods considering curve translations. First, the scaling that we proposed depends on the covariates (\emph{i.e.,} the risk factors). Second, we use a wavelet approach that permits a greater reduction of the dimension. Finally, we consider a semi-parametric mixture that avoids the bias of the parametric mixtures observed when their parametric assumptions are violated.

The remainder of this article is organized as follows. Section~2 presents the data. Section~3 defines the new approach \approach. Section~4 illustrates, with numerical experiments, the relevance of \approach. In Section~5, we analyze the geographical disparities of the \cov by investigating the regression coefficients, describing the clusters according to the diseases and illustrating the use of the clusters to investigate policy strategies. Section~6 presents some concluding remarks.

\section{Description of the data}   \label{sec:datadesc}
For this ongoing \cov dataset, we consider $n=94$ regions composed of the 50 states of the USA plus the District of Columbia and countries from Europe and America (Austria, Belgium, Bulgaria, Brazil, Canada, Switzerland, Chile, Colombia, Costa Rica, Cuba, Cyprus, Czechia, Germany, Denmark, Dominican Republic, Estonia, Spain, France, United Kingdom, Greece, Guyana, Croatia, Haiti, Hungary, Ireland, Iceland, Italy, Jamaica, Latvia, Netherlands, Norway, Panama, Peru, Poland, Portugal, Paraguay, Romania, Serbia, Sweden, Slovenia, Slovakia, El Salvador and Uruguay) whose data are available at Johns Hopkins' Github repository (file Policy.rds in \citet{Hopkins}). 

\begin{figure}[!htp]
\centering
\includegraphics[scale=0.35]{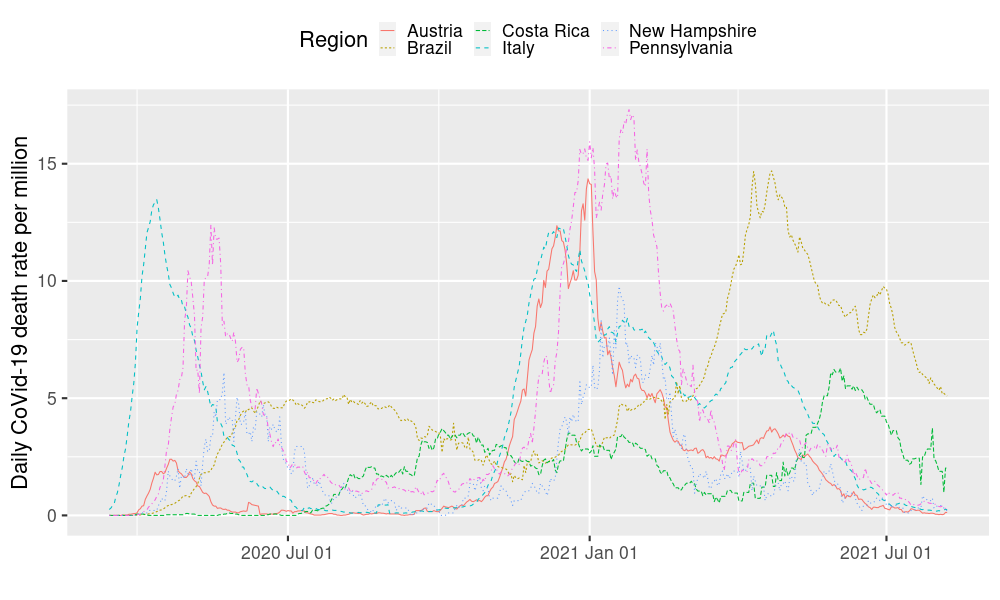}
\caption{Illustration of different arrivals of the \cov among the different regions (Europe, North America, Latina America).}\label{fig:realtrans}
\end{figure}

Our focus is on the curve $W_i=(W_{i(1)},\ldots,W_{i(T)})^\top$ recording the daily rate of the number of deaths per million people in each region $i$ for a total of $T=512$ days (between March 1st, 2020 to July 25, 2021, inclusively), where $W_{i(t)}$ denotes the death rate recorded for region $i$ at time $t$. Data were extracted from the Center for Systems Science and Engineering at Johns Hopkins' Github repository (file Policy.rds in \citet{Hopkins})  and a 7-day moving average has been performed due to the discrepancy of the data recorded by each region. For instance, this can account for days in the week where data may not be available, such as weekends. 
There are differences in time of arrival of the peak death rates between regions. This is illustrated by Figure~\ref{fig:realtrans} which shows that the first two COVID death peaks arrived first in Europe (see Austria and Italy), then in North America (see New Hampshire and Pennsylvania) and then in Latina America (Brazil and Costa Rica).

Early findings suggested that differences in \cov disease prevalence and severity may be associated with certain risk factors \citep{Williamson2020}. Thus, we consider two groups of risk factors. The first group contains three environmental risk factors (fine particulate matter PM2.5 concentration, nitrogen dioxide NO2 concentration  and population density) and the second group contains eight medical risk factors (age-adjusted percent prevalence of adults with diagnosed diabetes, percent of obese adults, age-adjusted percent prevalence of adults who are current smokers, age-standardized percent prevalence of chronic obstructive pulmonary disease, age-standardized percent prevalence of cardiovascular disease, age-standardized percent prevalence of HIV/AIDS, percent of adults with hypertension and population proportion over 65 years old). All the risk factors were extracted from the Center for Systems Science and Engineering at Johns Hopkins' Github repository (file COVID-19\_Static.rds in \citet{Hopkins}) and have been scaled. For each group of risk factors, we perform a principal component analysis (PCA). For each region $i$, we store in $X_i$ the first principal components of both groups of risk factors. In the application described in Section~\ref{sec:appli}, we show that it is relevant to consider the first principal component of the PCA based on the environmental risk factors and to consider the first two  principal components of the PCA based on the environmental risk factors. 
Note that that we are able to consider the states of the USA because all the data (risk factors and daily rate of the number of deaths) are available at Johns Hopkins' Github repository, however, these data were not available for the different provinces of Canada or the states of Brazil.

\section{Method}
\subsection{Outline of the three-step method}
The daily \cov death curves of the $n$ regions $W_1,\ldots,W_n$ are supposed to independently arise from $L$ different clusters. The cluster membership of region $i$ is defined by the latent variable $Z_i=(Z_{i1},\ldots,Z_{iL})^\top$ where $Z_{i\ell}=1$ if region $i$ belongs to cluster $\ell$ and $Z_{i\ell}=0$ otherwise. The model assumes that, conditionally on the cluster $\ell$, each $W_i$ is defined as a product between a noisy version of $\delta_i$-lagged values of an unobserved curve $u_\ell$ and the effect of the population risk characteristics $\mu(X_i)>0$, where $X_i$ denotes population risk factors of region $i$ and where we set $\mathbb{E}[\mu(X_i)]=1$, for identifiability reasons. The deterministic functions $u_{\ell}$ do not depend on the covariates $X_i$. Moreover the noises $\varepsilon_{i\ell}$ and the covariates $X_i$ are independent. Thus, given the cluster membership and the population risk factors of region $i$, we have
\begin{equation} \label{eq:modelcond} 
W_{i} = \sum_{\ell=1}^L z_{i\ell} \mu(x_i)(u_{\ell}^{(\delta_i)} + \varepsilon_{i\ell}^{(\delta_i)}),
\end{equation}
where $u_{\ell}^{(\delta_i)}$ and $\varepsilon_{i\ell}^{(\delta_i)}$ are $\delta_i$-lagged versions of $u_{\ell}$ and $\varepsilon_{i\ell}$, and the distribution of each $\varepsilon_{i\ell}$ follows a centered distribution having a finite variance defined by the density $f_\ell$ (\emph{i.e.}, $\mathbb{E}_{f_\ell}[\varepsilon_{i\ell}]=0$ and $\mathbb{E}_{f_\ell}[\varepsilon_{i\ell}^2]<\infty$).  Thus,  noting that $f(w_i\mid X_i=x_i)=\sum_{\ell=1}^L f(w_i\mid X_i=x_i, Z_{i\ell}=1)$,  $f(w_i\mid X_i=x_i, Z_{i\ell}=1)=\mathbb{P}(Z_{i\ell}=1\mid X_i=x_i)f(\varepsilon_{i\ell}\mid X_i=x_i, Z_{i\ell}=1)=f_\ell(\frac{w_i}{\mu(x_i)} - u_\ell^{(\delta_i)})$ and, by assumption $\mathbb{P}(Z_{i\ell}=1\mid X_i=x_i)=\mathbb{P}(Z_{i\ell}=1)=\pi_\ell $,  the conditional distribution of $W_i$ given $X_i=x_i$ is defined by the density
\begin{equation} \label{eq:modelstart}
f(w_i \mid x_i) = \sum_{\ell=1}^L \pi_\ell f_\ell\left( \frac{w_i}{\mu(x_i)} - u_\ell^{(\delta_i)} \right),
\end{equation}
where $\pi_\ell>0$ is the proportion of cluster $\ell$ with $\sum_{\ell=1}^L\pi_{\ell}=1$. 
Note that the model defined by \eqref{eq:modelstart} is related to the translation invariant model proposed by \citet{casa2021co} for co-clustering of multivariate functional data. However, in \cite{casa2021co} the time of translation $\delta_i$ is random variable that is relevant for clustering (\emph{i.e.,} the conditional distributions of $\delta_i$ given the cluster memberships are not equal for all the clusters) while this variable is supposed to be irrelevant for clustering in \eqref{eq:modelstart}. In addition, in \citet{casa2021co} the noises are supposed to be independent and identically distributed within a cluster and thus their distributions are not impacted by the time translation. In \eqref{eq:modelstart}, the noises are not supposed to independent and identically distributed within components and thus their distributions can be impacted by the time translation.

Despite the model defined by \eqref{eq:modelstart} permitting a clustering of the regions based on the daily \cov death curves  with respect to the population risk factors, the estimation the multivariate density $f_\ell$ is highly complex. Thus, we achieve the clustering with the following three-step approach:
\begin{enumerate}
\item Perform feature extraction of the daily \cov death curves $W_i$ to obtain $Y_i \in \mathbb{R}^{J+1}$ using \TI wavelets (see Section~\ref{sec:wavelets}).
\item Fit single-index regressions of the features $Y_i $ on the population risk factors $X_i$ and consider the residuals $\hat\xi_i\in\mathbb{R}^{J+1}$ (see Section~\ref{sec:SI}).
\item Use the nonparametric mixture to cluster the regions based on the residuals $\hat\xi_i$ (see Section~\ref{sec:clustering}).
\end{enumerate}
This approach is relevant since the specific feature extraction reduces the dimension, permits us to deal with lagged values and keeps the main cluster information. Moreover, the single-index regression keeps the cluster information of the features, allows for adjustment on population risk factors, and provides meaningful parameters used for detecting protective or compounding effects of the population characteristics and odd ratios.
 
\subsection{Feature extraction and time misalignment}
\label{sec:wavelets}
A wavelet basis is a set of functions obtained as translations and dilatations of two specific functions: a scaling function denoted by $\phi$ and a mother wavelet denoted by $\psi$. For the purpose of this paper, we use Daubechies wavelets and in particular the Symlet family. We present the essentials below, more details can be found in \cite{daub} or \cite{mallat:08}.  Such wavelets are optimal in the sense that they have minimal support for a given number of null moments. For the simulations, we consider in particular the Symlet family (or Daubechies least asymmetric) which is a modified version, with more symmetry than the classical Daubechies wavelets. We present the essentials below, more details can be found in \cite{daub} or \cite{mallat:08}.  

The scaling function $\phi$ and wavelet function $\psi$ of an orthonormal Daubechies wavelet system satisfy the scaling equations
\[
\phi(t) =\sqrt{2} \sum_{k\in\mathbb{Z}}h_k\phi(2t-k), \qquad \psi(t) = \sqrt{2} \sum_{k\in\mathbb{Z}}g_k\phi(2t-k),
\]
with scaling and wavelet filters $\bm h=\{h_k:k\in \mathbb{Z}\}$ and $\bm g=\{g_k:k\in\mathbb{Z}\}$, with $g_k=(-1)^kh_{1-k}$. For any $j\ge 0$, we set $\Lambda_j=\{0,\ldots,2^j-1\}$ and, for $k\in \Lambda_j$,
\[
\phi_{j,k}(t)=2^{j/2}\phi(2^jt-k),  \qquad \psi_{j,k}(t)=2^{j/2}\psi(2^jt-k).
\]
Following the methodology of \cite{cohen}, there exists an integer $\tau$ such that, for any integer $j_0\ge \tau$, the collection of functions 
\[
\mathcal{S}=\{ \phi_{j_0,k}, \ k\in \Lambda_{j_0}; \ \psi_{j,k}; \ j \in \mathbb{N}-\{0,\ldots, j_0-1\} ,\ k\in \Lambda_{j}\}
\]
forms an orthonormal basis of $\mathbb{L}^2(\lbrack 0,1 \rbrack)$. Therefore, for any integer $j_0 \ge \tau$ and $W\in \mathbb{L}^2(\lbrack 0,1 \rbrack)$, we have the following wavelet expansion
\begin{equation}
W(t)= \sum_{k\in \Lambda_{j_0}}\alpha_{j_0,k}\phi_{j_0,k}(t)+\sum_{j=j_0}^{\infty}  \sum_{ k\in\Lambda_{j}} \beta_{j,k}\psi_{j,k}(t), \quad t \in [0,1],
\end{equation}
where $\alpha_{j,k}$ and $\beta_{j,k}$ are the scaling and wavelet (or details) coefficients of $f$ at scale $j$ and position $k$ defined as
\begin{equation*}
\alpha_{j_0,k}=\int_{0}^1f(t)\phi_{j_0,k}(t)dt, \quad \beta_{j,k}=\int_{0}^1f(t)\psi_{j,k}(t)dt.
\label{coef}
\end{equation*}
The first term of this development can be seen as an approximation term of the function at level $j_0$ and the second as a detail term that characterizes the approximation error. The level of approximation $j_0$ has relatively no influence on the performance of the approximation (or estimation) and will subsequently be set to $0$. 
The decomposition of the observations in a given wavelet basis is defined by
\begin{equation*}
W_i(t)= \alpha_{i,0,0}\phi_{0,0}(t)+\sum_{j=0}^{J-1}  \sum_{k=0}^{2^j-1} \beta_{i,j,k}\psi_{j,k}(t),\quad t \in [0,1],
\end{equation*}
with $J=\log_2(T)$, $\alpha_{i,0,0}\approx \sqrt{T}\int_{0}^1W_i(t)\phi_{0,0}(t)dt$ and $\beta_{i,j,k}\approx\sqrt{T}\int_{0}^1W_i(t)\psi_{j,k}(t)dt$ are the empirical wavelet coefficients of the $i$th individual.
A discrete wavelet transform (DWT) corresponds to the computation of these coefficients. 
In practice, a fast wavelet decomposition and reconstruction algorithm can be computed using the algorithm proposed by \cite{mallat1989theory} (in only $\mathcal{O}(T)$ operations). 
As mentioned in the introduction, a simple shift in the observed function will potentially result in a significant change in the DWT. Since we use the latter for feature extraction and the observed curves can start at different times, such behavior is not suitable. 
 
In the \TI case, we consider the fast translation-invariant discrete wavelet transform (TIDWT) developed by \cite{Coifman1995}, in a denoising framework. This transformation has been independently discovered, on several occasions, in different communities, and has received several different names, including the  ``\textit{{\`a} trous}" algorithm \cite{holschneider1990real,dutilleux1990implementation}, the undecimated DWT \cite{lang1996noise}, the shift-invariant DWT \cite{lang1995nonlinear} or the stationary DWT \cite{nason1995stationary}, to name just a few (see, \emph{e.g.} \cite{fowler2005redundant} for a review of some of the various different variants). There are many ways to implement this transformation, and many ways to represent the resulting overcomplete set of wavelet coefficients. We have chosen to focus on the TIDWT of \cite{Coifman1995}, which provide equivalences and a way to go from one to the other of these representations, for example with the stationary DWT of \cite{nason1995stationary}.
The main difference with the orthogonal case is that the dictionary is now a tight frame instead of an orthonormal basis (see \cite[Chapter 5]{mallat:08}) and the number of coefficients per scale is no longer dyadic but of length $T$ (see \cite{Coifman1995} for more details). This wavelet transform is called translation-invariant by \cite{Coifman1995} since the whole dictionary is invariant under circular translation. 
As in the traditional case, TIDWT is calculated by a series of decimation and filtering operations, only the additional circulant shift $S_h$ is added and the corresponding wavelet dictionary is obtained by sampling the locations more finely (\emph{i.e.}, one location per sample point). 
TIDWT consists of calculating the DWT of the shifted data for each shift $h\in\{0,\ldots,T-1\}$. \cite{Coifman1995} propose an algorithm to perform this transformation in $\mathcal{O}(T\log_2 T)$ operations (we used the \proglang{R} package \pkg{rwavelet} which provides an implementation \cite{navarro2020r}). The invariance property of their construction is formally expressed in terms of the circulant matrix containing the wavelet coefficients (see \cite[eq. (3)]{Coifman1995}). In other words, for a curve $W_i$ translated by $h$ the wavelet coefficients at each scale will be the same up to some permutation. Thus the norm of the latter is preserved scale by scale.

The redundancy of TIDWT makes it possible to detect the presence of hidden information such as stationary or non-stationary patterns as well as their location, making it particularly suitable for clustering purposes. This type of invariant representation has been exploited in many applications (such as denoising \cite{Coifman1995} or texture image classification and segmentation \cite{unser1995texture}). In addition, the use of wavelets allows to compress the information contained in the time series into a small number of wavelet coefficients. Following \cite{antoniadis2013clustering}, we characterize each time series by the vector of the energy contribution of their wavelet coefficients at each scale with the difference that the coefficients are calculated by TIDWT instead of DWT.
This extension is possible because the expansion being in a \textit{tight frame}, the norm is also conserved (see \cite[Chapter 11]{mallat:08} for more details). More precisely, using Parseval's identity, we have
\begin{equation}\label{parseval}
\|W_i\|^2_2=2^{-J}\sum_{k=0}^{T-1}\alpha_{i,0,k}^2+\sum_{j=1}^{J}{2^{-j}}\sum_{k=0}^{T-1} \beta_{i,j,k}^2=2^{-J}\|\theta_{i0}\|^2_2+\sum_{j=1}^{J}{2^{-j}}\|\theta_{ij}\|_2^2,
\end{equation}
where $\theta_{ij}=(\alpha_{i,0,0},\ldots,\alpha_{i,0,T-1},\beta_{i,j,0},\ldots,\beta_{i,j,T-1})^\top$ and the factor $2^{-j}$ is used to compensate for the redundancy of this representation.
The global energy $\|W_i\|^2_2$ of $W_i$ is decomposed into a small number of components. The representation \eqref{parseval} is made of the components of the invariant version of the discrete wavelet scalogram (as defined in \cite{arino2004wavelet}) can be considered as the TIDWT analogue of the well-known periodogram of the spectral analysis of time series. Similar to how the periodogram produces an ANOVA decomposition of a signal's energy into different Fourier frequencies, the scalogram breaks down the energy into ``level components" that indicate at which levels of resolution the energy of the observed function is concentrated. A function that is relatively smooth will have most of its energy concentrated in the low levels $j$, resulting in a $\theta_{ij}$ that is large for small $j$ and small for large $j$.  A function with many high-frequency oscillations will have much of its energy concentrated in the high-resolution wavelet coefficients. Therefore, how these energy components are distributed and contribute to the overall energy of a signal is the key fact that we will exploit for clustering. 
Thus, denoting by $y_{ij}$ the log total squared norm at scale $j$ for the $i$th individual, we have 
\begin{equation}\label{energy}
y_{ij} = \ln\left(\|\theta_{ij}\|_2^2\right), \quad \forall  j=0,\ldots,J,\quad i=1,\ldots,n.
\end{equation}
Clustering will therefore be carried out on the basis of the log squared norm of the \TI wavelet coefficients at each scale. Thus, this criterion is not sensitive to the origin of the curves, so it seems relevant given the nature of the data motivating this work.

\subsection{Adjustment on the population risk factors} 
\label{sec:SI}
In this section, we consider the regressions of the features extracted by the wavelet decomposition on the population risk factors. The following lemma shows that the noises of these regressions retain the cluster information given by the daily \cov death curves and permit the information of the  population risk factors to be considered in the clustering procedure. 
Thus, the effects of the risk factors can be estimated from the wavelets coefficients with a nonparametric regression (see \eqref{eq:SImodel}). Then, the cluster memberships of the region can be assessed from the residuals of the regressions (see \eqref{eq:mixtureresiduals}) since this residuals keep the discriminative information contained in the original data. 
Note that the same nonparametric function is used for the regression of each feature (\emph{i.e.}, $j=0,\ldots,J$).

\begin{lemma}\label{lem:inforesiduals}
Let data $W_1,\ldots,W_n$ arise from \eqref{eq:modelstart} and $J+1$ features  $y_{i1},\ldots,y_{iJ}$ are defined by \eqref{energy} from the wavelet decomposition of $W_i$. Let the centered features be defined by $y_{ij}^\star=y_{ij} - \Delta_j$,  $\Delta_j=\mathbb{E}[\ln \mu(X_i)]+ \sum_{\ell=1}^L \pi_\ell \mathbb{E}[\frac{1}{2} \ln \|v_{\ell j} + \varepsilon_{i\ell j}^\star\|_2^2]$, and $v_{\ell j}$ and $\varepsilon_{i\ell j}^\star$  are computed from $u_\ell$ and $\varepsilon_{i\ell}$, respectively (their formal definition is given in the proof of the lemma presented in Appendix).
\begin{equation}\label{eq:SImodel}
y_{ij}^\star = m(x_i) + \xi_{ij},
\end{equation}
with $m(x_i) = \ln \mu(x_i) - \mathbb{E}[\ln \mu(X_i)]$,
$$\mathbb{E}[m(X_i)]=0 \text{ and } \mathbb{E}[\xi_{ij}]=0,$$
where the covariates $x_i$ and the noise of the regression $\xi_{ij}$ are independent,for $j=0,\ldots,J$.  Then, the noises $\xi_i$ follow a mixture model with latent variable $Z_i$ defined by the density
\begin{equation} \label{eq:mixtureresiduals}
g(\xi_i)=\sum_{\ell = 1}^L \pi_\ell g_\ell(\xi_i - \lambda_\ell ),
\end{equation}
where $\lambda_\ell=(\lambda_{\ell 1},\ldots,\lambda_{\ell J})^\top$, $\lambda_{\ell j}= \mathbb{E}[\frac{1}{2} \ln \|v_{\ell j} + \varepsilon_{i\ell j}^\star\|_2^2] - \Delta_j + \mathbb{E}[\ln \mu(X_i)]$ and $g_1,\ldots,g_\ell$ are densities of centered distributions.
\end{lemma}

We consider the single-index regression defined by 
\begin{equation}\label{eq:SI}
m(x_i):=\nu(x_i^\top\gamma).
\end{equation}
This semiparametric approach is flexible, and avoids the assumptions of the parametric approaches that can be violated and the curse of dimensionality of the full nonparametric approaches. The parameter of the index $\gamma$ permits population characteristics having a protective or compounding effect to be detected. Moreover, considering two sets of covariates $X_i$ and $X_{i'}$, the difference $\nu(X_i^\top \gamma)-\nu(X_{i'}^\top \gamma)$ can be interpreted as the logarithm of an odd ratio. 
 
The single-index approach requires a methodology for estimating $\gamma$ and $m$, with $m$ being in a function space.  A common approach, that avoids a simultaneous search involving an infinite-dimensional parameter, is the profiling \citep{severini1992,liang2010}, which defines $\nu(x_i^\top\gamma):=\nu_{\gamma}$ with
\begin{equation}\label{eq:mfct}
\nu_{\gamma} (t) = \mathbb{E}[Y_{ij}^\star \mid X_i^\top \gamma=t],\qquad j\in\{0,\ldots,J\} \text{ and } t\in\mathbb{R}. 
\end{equation}
Hence, one expects that, for each $x_i$, the true value of the parameter, denoted by $\gamma$ realizes  the minimum of 
\begin{equation}\label{eq:loss} 
\gamma\mapsto \sum_{j=0}^J \mathbb{E}[\{ Y_{ij}^\star- \nu_{\gamma}(x_i^\top \gamma) \}^2\mid X_i=x_i].
\end{equation}
However, even if $m_{\gamma}$ is well defined for any $\gamma \in \mathbb R^d$, the vector $\gamma$ is not identifiable and only its direction could be consistently estimated. Thus, there are two common approaches to restrict $\gamma$ for identification purposes: either fix one component equal to 1 \citep{Ma2013}, or set the norm of $\gamma$ equal to 1 and fix the sign of its first component to be positive \citep{zhu2006empirical}. The estimation of the single-index regressions is performed by considering the empirical counterpart of \eqref{eq:mfct} and \eqref{eq:loss}
\[
\hat\gamma = \argmin_{\gamma} \sum_{i=1}^n\sum_{j=1}^J \left(\hat{y}^\star_{ij} - \hat \nu_{\gamma}(X_{i}^\top \gamma)\right)^2,
\]
\[
\hat{y}^\star_{ij} = y_{ij} - \frac{1}{n}\sum_{i=1}^n y_{ij},\quad \textrm{and} \quad \hat \nu_{\gamma} (u) =  \frac{
\frac{1}{nh} \sum_{i=1}^n \hat{y}^\star_{ij} K\left(\frac{X_i^\top \gamma - u}{h}\right)
}{
\frac{1}{nh} \sum_{i=1}^n K\left(\frac{X_i^\top \gamma - u}{h}\right)
},
\]
where $K$ is a kernel and $h$ a bandwidth. The estimation procedure is implemented in the \proglang{R} package \pkg{regpro} \citep{regpro}. 
The clustering of the regions is also performed on the residuals $\hat \xi_i=(\hat \xi_{i0},\ldots,\hat \xi_{iJ})^\top$ defined by
\begin{equation*}
\hat \xi_{ij} = \hat{y}^\star_{ij} - \hat \nu_{\hat\gamma}(X_{i}^\top \hat\gamma). \label{eq:residuals}
\end{equation*}

\subsection{Nonparametric clustering of the regions}
\label{sec:clustering}
A wide range of literature focuses on models assuming that, conditionally on knowing the particular cluster the subject $i$ came from, its  features are independent \citep{chauveauSurveys2015}. Thus, we consider that the conditional distribution of the $\hat \xi_{i}$ given cluster membership  is defined as a product of univariate densities. 
Note that this assumption is standard when nonparametric modelling of the mixture components is wanted. Moreover, this assumption is suitable when $J$ is quite large with respect to $n$ because it limits the number of parameters to be considered.
Note that this assumption imposes non-explicit constraints on the distribution of the noises $\varepsilon_i$ defined in \eqref{eq:modelcond}.
Note that this assumption could be relaxed by in a parametric framework \citep{bouveyron2015funfem} but also in a nonparametric framework \citep{zhu2019clustering}. In numerical experiments (see Section~4.1), we illustrate the relevance of the conditional independence assumption.
Therefore, the clustering of the region is performed by considering the marginal density defined by
\begin{equation} \label{eq:mixture}
g(\hat \xi_{i};\lambda) = \sum_{\ell=1}^L \pi_\ell \prod_{j=1}^J g_{\ell j}(\hat \xi_{ij}),
\end{equation}
where $\lambda$ groups the mixing proportions $\pi_1,\ldots,\pi_L$ (where $\pi_\ell >0$ and $\sum_{\ell=1}^L \pi_\ell=1$) and the univariate densities $g_{\ell j}$. The model \eqref{eq:mixture} is identifiable, up to a swapping of the cluster labelling, if the densities $g_{\ell j}$ are linearly independent (see Theorem~8 of \cite{allman2009identifiability}).
 Considering a multivariate kernel defined as a product of $J$ univariate kernels $K$, the maximum smoothed log-likelihood estimator $\hat\lambda$ (MSLE) is obtained by maximizing the smoothed log-likelihood $\ell(\lambda)$ \citep{levine2011maximum}, such that
\begin{equation*}
\hat\lambda_L = \argmax_\lambda \ell(\lambda;L)
\end{equation*}
and
\begin{equation*}
\ell(\lambda;L)= \sum_{i=1}^n \ln\left\{ \sum_{\ell =1}^L \pi_\ell \prod_{j=1}^J \mathcal{N} g_{\ell j}(\hat \xi_{ij})\right\},
\end{equation*}
where
\begin{equation*}
\mathcal{N}g_{\ell j}(\hat \xi_{ij})=\exp\left\{\int_{\Omega_j} \frac{1}{h_j} K\left(\frac{\hat \xi_{ij} - u}{h_j}\right)\ln g_{\ell j}(u)du\right\},
\end{equation*}
and $h_1,\ldots,h_J$ are the bandwidths (\emph{i.e.}, $h_j>0$ and $h_j=o(1)$ for $j=1,\ldots,J$). Considering the MSLE is more convenient than considering the maximum likelihood estimate because the MSLE can be obtained by a Majorization-Minimization algorithm (see \cite{levine2011maximum} for details on the algorithm and \cite{ZhuJNPS2016} for recent developments) implemented in the \proglang{R} package \pkg{mixtools} \cite{benaglia2009mixtools}.

Clustering is achieved by computing the MSLE because this estimator permits a soft assignment where the conditional probability that subject $i$ belongs to cluster $\ell $, denoted by $t_{i \ell}(\hat\lambda)$, can be obtained
\begin{equation*}
t_{i\ell }(\hat\lambda_L) = \frac{\hat\pi_\ell \prod_{j=1}^J \mathcal{N} \hat g_{\ell j}(\hat \xi_{ij})}{\sum_{\ell' =1}^L \hat\pi_{\ell'} \prod_{j=1}^J \mathcal{N} \hat g_{\ell' j}(\hat \xi_{ij})}.
\end{equation*}
Moreover, a hard assignment can be achieved by applying the maximum \emph{a posteriori} rule (leading that $\hat{z}_{i\ell}=1$ if $\ell =\argmax_{\ell'} t_{i\ell }(\hat\lambda)$ and  $\hat{z}_{i\ell}=0$ otherwise).

\section{Numerical experiments}
\subsection{Investigating the strengths of the proposed approach}
Data are independently generated from \eqref{eq:modelcond} with $T=512$, $K=3$, unequal proportions ($\pi_{1}=0.5$, $\pi_2=0.25$ and $\pi_3=0.25$) by considering two scenarios:
\begin{itemize}
\item For scenario 1, we define $$u_{\ell (t)}^{(\delta_i)}= \left\{\begin{array}{rl}
r_{\ell}(t-\delta_i) \mathds{1}_{\{r_{\ell}(t-\delta_i) >0\}} & \text{ if } t>\delta_i\\
0 & \text{otherwise} \end{array}\right.,$$ with $r_{1}(t) =\sin(2.5 \pi t/T)$,  $r_{2}(t) =1.5 \sin(2.5 \pi t/T)$ and $r_{3}(t) =\sin(2 \pi t/T)$ and the vector of noises $\varepsilon_{i\ell}$ is composed of independent Gaussian random variables with mean 0 and variance $1/2$.
\item For scenario 2, we define $$u_{\ell (t)}^{(\delta_i)}= \left\{\begin{array}{rl}
r_{\ell}(t-\delta_i)  & \text{ if } t>\delta_i\\
0 & \text{otherwise} \end{array}\right.,$$ 
and consider the same simulation setting presented in \cite[Section 4.1]{antoniadis2001wavelet}. That is,  $r_{1}(t)=\sin(2 \pi t/T)+\sin(5 \pi t/T)$ and for the second and third clusters, we use two strictly stationary autoregressive Hilbertian (ARH) processes with the same parameters (see \cite[Section 4.1]{antoniadis2001wavelet} for more details).
\end{itemize}  
For both scenarios, the time shift $\delta_i$ of curve $i$ is generated from a uniform distribution defined on $\{0,\ldots,\Delta\}$. Moreover, we generate two covariates from standard Gaussian distributions independently, and we consider $\gamma=(1/\sqrt{2},1/\sqrt{2})^\top$ and $\nu(t)=(1+q t^2)/(1+q) $. 

\begin{figure}[htp!]
\begin{center}
\includegraphics[scale=0.4]{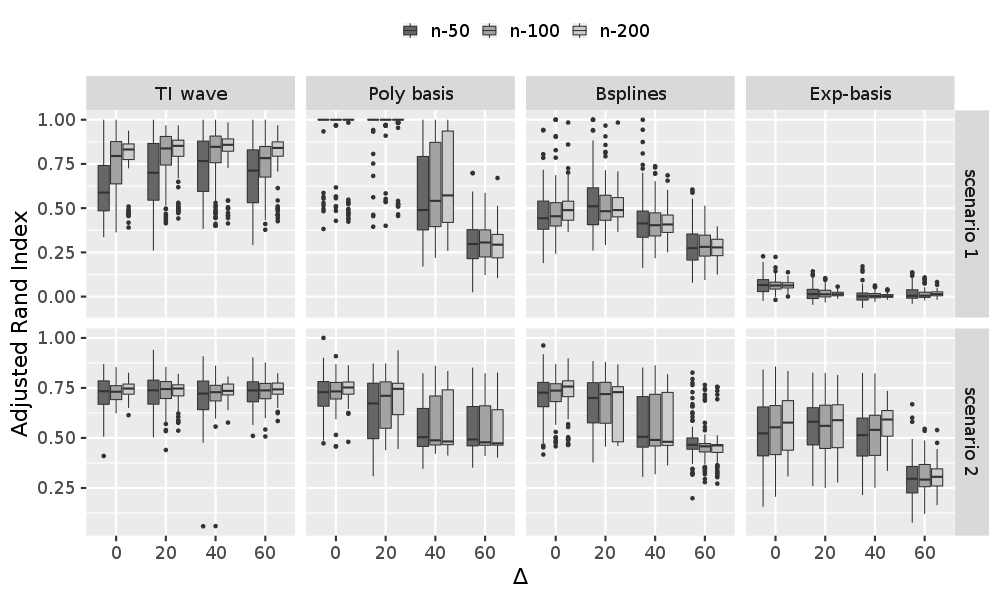}
\caption{Boxplots of the Adjusted Rand Index obtained on 100 replicates according to the basis decomposition (TI-wave,  Poly-basis, Bspline and Exp-basis), the sample size (50, 100 and 200) and the value of $\Delta$ (0, 20, 40 and 60).}\label{fig:simuTI}
\end{center}
\end{figure}

\paragraph{Relevance of the translation-invariant feature extraction}
This experiment illustrates the importance of using translation-invariant decomposition when the observed curves are time-shifted. Thus, data are generated with different values of $\Delta$ and  no covariate effects (\emph{i.e.,} $q=0$).  To investigate the relevance of the proposed approach, we compare the translation-invariant wavelet  decomposition (TI-wave) with different functional basis:  polynomial basis (Poly-basis), Bsplines (Bsplines) and exponential basis (Exp-basis). Each basis decomposition considers 9 basis elements. For each basis decomposition, the clustering is achieved by fitting the nonparametric mixture model defined by \eqref{eq:mixture} on the basis coefficients. Figure~\ref{fig:simuTI} shows the Adjusted Rand Index (ARI) obtained on 100 replicates by the different basis decomposition for different sample sizes and different values of $\Delta$, under both scenarios. When there is no time shifting (\emph{i.e.,} $\Delta=0$), under scenario 1 the polynomial basis outperforms all the other approaches, while  the translation-invariant wavelet basis, polynomial basis and the exponential basis perform similarly under scenario 2. Note that exponential basis with 9 elements seems not to be relevant for these scenarios. When the time-shifting increases, then all the results obtain by the different basis are (strongly) deteriorated expect those obtained by the translation-invariant wavelet frame. These results illustrate the importance of applying a translation-invariant clustering method to analyze the \cov data. In the next section, we compare the proposed approach with other translation-invariant approaches.

\begin{figure}[htp!]
\begin{center}
\includegraphics[scale=0.4]{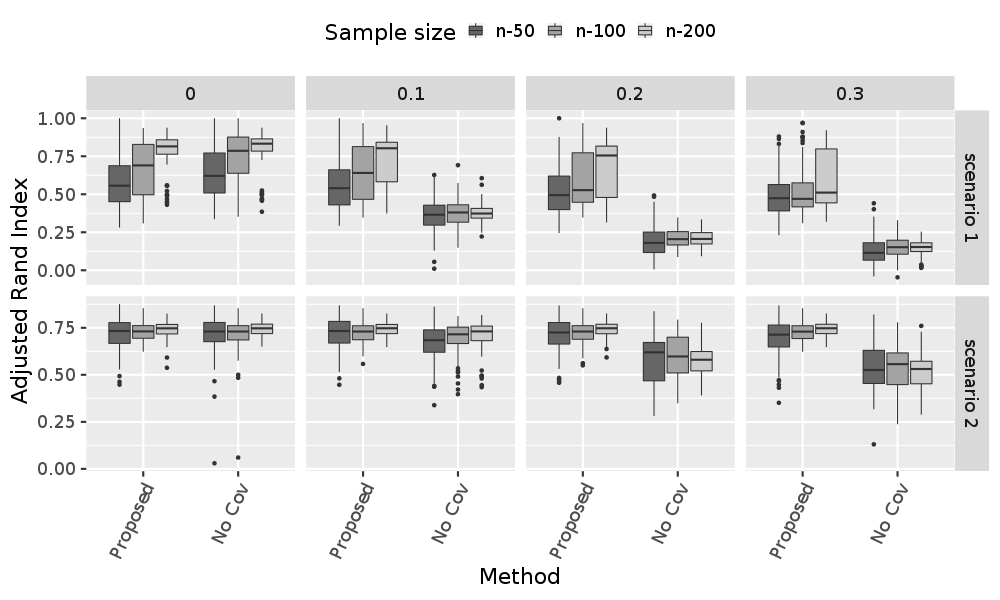}
\caption{Boxplots of the Adjusted Rand Index obtained on 100 replicates by considering (proposed) and by neglecting (No cov) the covariate effects when data are generated with different covariate effects $q$ (0, 0.1, 0.2 and 0.3).}\label{fig:simuCov}
\end{center}
\end{figure}

\paragraph{Relevance of the covariate adjustment}
This experiment illustrates the importance of considering the covariate effects. Thus, data are generated with  different covariate effects ($q\in\{0,0.1,0.2,0.3\}$) and no time shifting (\emph{i.e.,} $\Delta=0$).  To investigate the relevance of the proposed approach, we compare the results obtained by the proposed three-step approach with those obtained by neglecting the covariate effect (\emph{i.e.,} $\nu$ is supposed to be the constant function equals to one).
Figure~\ref{fig:simuCov} presents the distribution of the ARI obtained by the proposed approach that considers the covariate effect and by a clustering performed on the translation-invariant wavelet frame which neglects the covariate effect. When there is no covariate effect (\emph{i.e.,} $q=0$), the approach modeling the covariate effects obtained results that are slightly outperformed by the approach neglecting the covariate effects. However, when there is a covariate effect, the results obtained by the approach neglecting the covariate effects are strongly deteriorated while modeling the covariate effect allows the partition to be consistently estimated.

\begin{figure}[htp!]
\begin{center}
\includegraphics[scale=0.4]{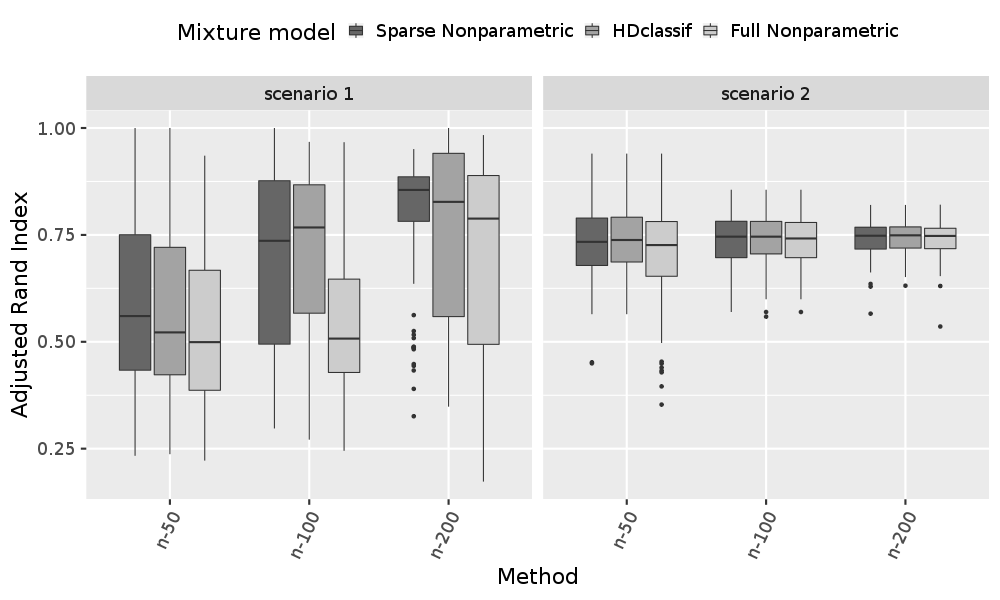}
\caption{Boxplots of the Adjusted Rand Index obtained by the different mixture models (Sparse non-parametric corresponds to the proposed approach, HDclassif and a Full nonparametric mixture) on 100 replicates according to the sample size.}\label{fig:simuClustering}
\end{center}
\end{figure}

\paragraph{Relevance of the nonparametric clustering}
This experiment focuses on the influence of the mixture model used for clustering.  Thus, data are generated with a time shift defined by $\Delta=20$ and a covariate effect defined by $q=0.1$. To investigate the relevance of the mixture model defined by \eqref{eq:mixture},  clustering of $\hat{\xi}_1,\ldots,\hat{\xi}_n$ is performed by considering \eqref{eq:mixture}, the semiparametric mixture considering the intra-component dependencies proposed by \citet{zhu2019clustering}, and the parsimonious Gaussian mixture models implemented in the \proglang{R} package \pkg{HDclassif} \citep{berge2012hdclassif}. Note that performing clustering on the basis coefficients with the Gaussian mixture implemented in \proglang{R} package \pkg{HDclassif} is the method proposed by \citet{bouveyron2015discriminative} and implemented in the \proglang{R} package \pkg{funFEM} \citep{bouveyron2015funfem}. Figure~\ref{fig:simuClustering} presents the distribution of the ARI according to the clustering method. Overall, the results of the three clustering methods are similar, despite the fact that the results obtained by considering intra-component dependencies slightly deteriorate the results of the nonparametric mixture. Under scenario 1, the proposed method (nonparametric mixture model with components defined as a product of univariate densities) slightly outperforms the parsimonious Gaussian mixture. Under scenario 2, all the methods perform similarly. These results show that the choice of the clustering method (step 3 of the proposed approach) is less impacting that the choice of the basis expansion (step 1 of the proposed approach) and the modeling of the covariate effect (step 2 of the proposed approach).

\subsection{Comparing the proposed approach with other methods used for \cov studies} \label{sec:simuCov}
\begin{figure}[htp!]
\begin{center}
\includegraphics[scale=0.45]{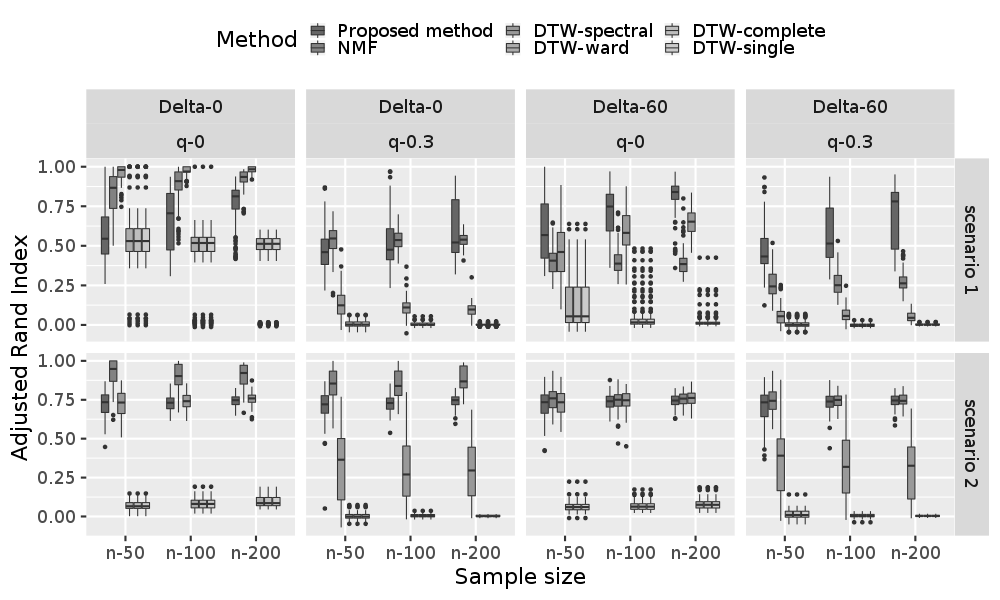}
\caption{Boxplots of the Adjusted Rand Index obtained on simulated data by the different clustering methods used in \cov studies.}\label{fig:compareCOVID}
\end{center}
\end{figure}

In this experiment, we compare the proposed three-step approach with alternatives used in other studies about \cov. Thus, data are generated from both scenarios described in the previous section with different values of $\Delta$ and $q$. Figure~\ref{fig:compareCOVID} presents the distribution of the ARI obtained by the proposed method, by the non-negative matrix factorization (NMF) approach \citep{chen2020clustering} and by using the similarity matrix given defined by the DTW approach for a clustering achieved via spectral clustering \citep{allem2020spectral} and via hierarchical ascendant classification \citep{park2020clustering} defined with different criteria (complete linkage, ward criterion, single linkage). When there is no time shifting and no covariate effect ($\Delta=0$ and $q=0$), NMF outperforms the proposed method under both scenarios and the spectral clustering performed on the DTW dissimilarity matrix outperforms the proposed method under scenario 1. However, our real data analysis considers regions from America and Europe, then time-shifting occurs. The results of the NMF methods are strongly deteriorated when data are time-shifted ($\Delta=60$) under both scenarios. Thus, NMF seems to be relevant to focus on a particular region where the diseases appear in the same time. Moreover, when there is a covariate effect, the results obtained by DTW-based approach deteriorated. To the best of our knowledge, there is no method that allows covariate to be included in DTW-based similarity matrix. Thus, the proposed method seems to be the single approach that allows to deal with time-shifting and to consider covariate adjustments.

\begin{figure}[htp!]
\begin{center}
\includegraphics[scale=0.45]{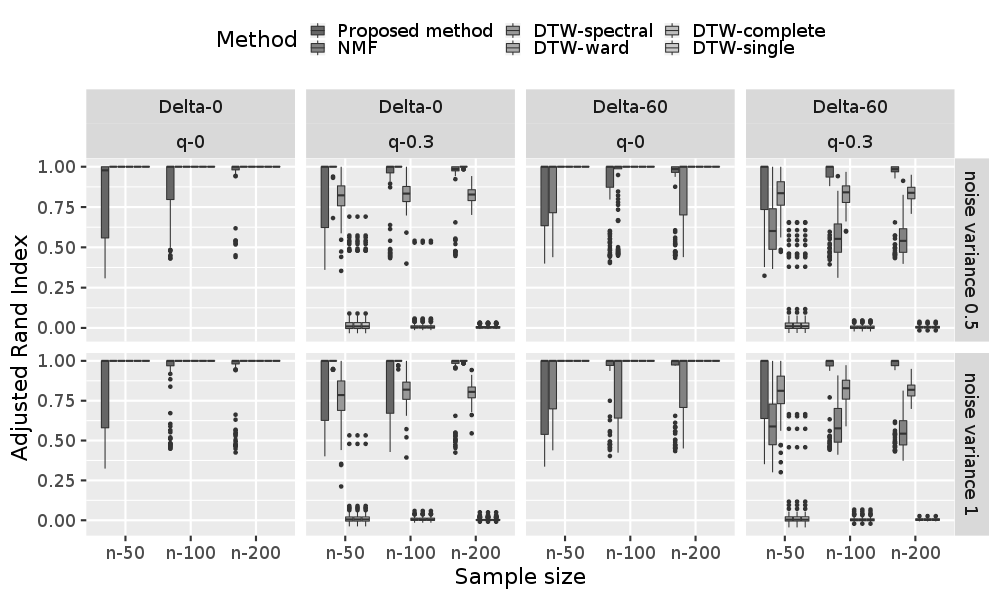}
\caption{Boxplots of the Adjusted Rand Index obtained on data simulated from the \cov dataset by the different clustering methods used in \cov studies.}\label{fig:compareCOVID2}
\end{center}
\end{figure}

We now use the \cov data to generate the samples. We consider the daily COVID death curves of Germany, Denmark and France (denoted by $\tilde{W}_1$, $\tilde{W}_2$ and $\tilde{W}_3$). Data are independently generated from \eqref{eq:modelcond} with $T=512$, $K=3$, equal proportions 
and 
$$
W_i(t)=\sum_{\ell=1}^L z_{i\ell} \mu(x_i^\top \gamma)(\tilde{W}_{\ell}^{(\delta_i)} + \varepsilon_{i\ell}^{(\delta_i)}).
$$
The time shift $\delta_i$ of curve $i$ is generated from a uniform distribution defined on $\{0,\ldots,\Delta\}$, the two covariates are generated from standard Gaussian distributions independently,  $\gamma=(1/\sqrt{2},1/\sqrt{2})^\top$, $\mu(t)=(1+q t^2)/(1+q) $ and the noises $\varepsilon_{i\ell}$ arise independently from a centered Gaussian distribution with variance $v^2$.  Figure~\ref{fig:compareCOVID2} presents the distribution of the ARI obtained by the competing methods. It shows that the proposed method outperforms the other competing methods when there is time shifting and a covariate effect.

\subsection{Investigating the robustness of the proposed approach}
We now investigate the robustness of the proposed approach to spatial dependencies. Indeed, these dependencies are not considered during the estimation but can occur in the \cov data set. Thus, we consider the two experiments described in Section~\ref{sec:simuCov} but relaxing the assumption of independence between the $\varepsilon_{i\ell}$. Thus, we now consider that the each vector $(\varepsilon_{1}(t),\ldots,\varepsilon_{n}(t))$ independently arises from a centered Gaussian random field with parameters $\sigma^2$ and $\phi$ where the distance between two sites is computed from their spatial coordinates that are generated uniformly in $[0,1]^2$.
Figure~\ref{fig:compareCOVID} presents the distribution of the ARI obtained  by the competing methods when the noises follow  Gaussian random fields.  Figure~\ref{fig:compareCOVID2b} presents the distribution of the ARI obtained by the competing methods when the data are simulated from the \cov dataset. Results show that the proposed method outperforms the other competing methods when there is time shifting and a covariate effect. Moreover, the spatial dependency does not deteriorate the results.

 \begin{figure}[htp!]
\begin{center}
\includegraphics[scale=0.45]{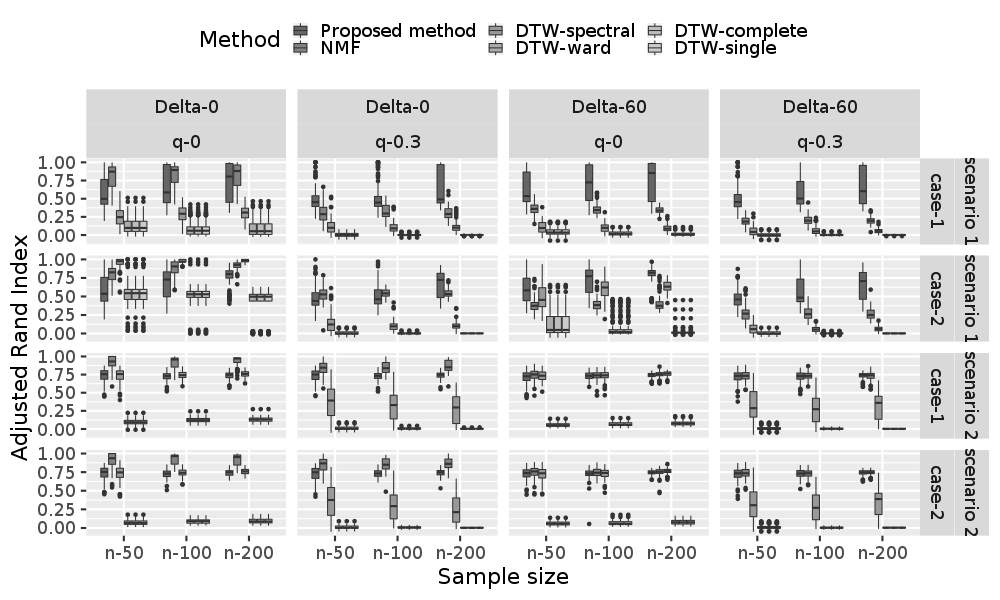}
\caption{Boxplots of the Adjusted Rand Index obtained on simulated data by the different clustering methods used in \cov studies when the noises follow  Gaussian random fields.}\label{fig:compareCOVIDb}
\end{center}
\end{figure}

\begin{figure}[htp!]
\begin{center}
\includegraphics[scale=0.45]{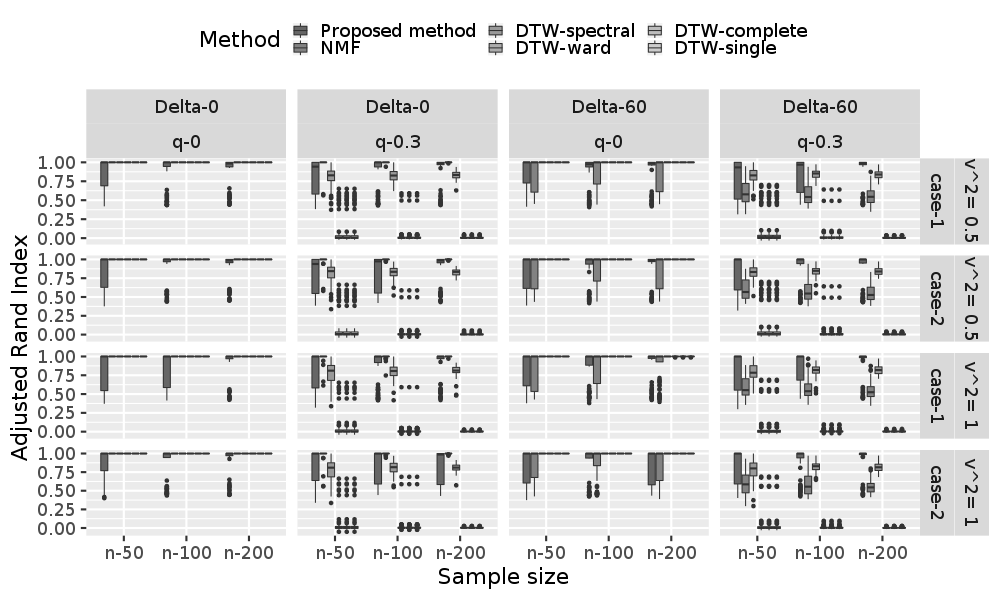}
\caption{Boxplots of the Adjusted Rand Index obtained on data simulated from the \cov dataset by the different clustering methods used in \cov studies when the noises follow  Gaussian random fields.}\label{fig:compareCOVID2b}
\end{center}
\end{figure}

\section{Investigating geographical disparities for \cov} \label{sec:appli}
The proposed approach allows risk factors to be taken into account, and thus to investigate the geographical disparities of the \cov that are not explained by geographical disparities of the population risk factors. Thus, this adjustment permits similarities and disparities of the disease impacts to be detected, conditional on the population characteristics. In  Section~\ref{appli:risk}, we interpret the impact of the population risk factors on the \cov death curves and we confirm this interpretation with literature review. In Section~\ref{appli:cluster}, we describe the clustering outputs. Finally, Section~\ref{appli:supp} presents an \emph{a posteriori} comparison of the clusters with respect to different policy decisions (\emph{e.g.}, lockdown characteristics).

\subsection{Population risk factors} 
\label{appli:risk}
During the application, the first principal component obtained on the PCA of the environmental variables and the first two principal components obtained on the PCA of the medical variables can be considered. Indeed, the second principal component obtained on the PCA of the environmental variables and the third principal component obtained on the PCA of the medical variables are not relevant in the single-index regression (see \eqref{eq:SI}). The relevance of the estimated coefficients of the single-index regression $\widehat{\gamma}$ is investigated by using empirical likelihood \citep{dechaumaray2020wilks}. 
Table~\ref{tab:EL} presents the estimator of the regression coefficients and its p-value provided by using the empirical likelihood ratio. It also presents the estimators of the coefficients obtained by assuming that one of the four variables does not have an effect on the target variables (where $\widehat{\gamma}^{\{\omega\}}=\sum_{i=1}^n\sum_{j=1}^J \left(\hat{y}^\star_{ij} - \hat \nu_{\gamma}(X_{i}^\top \gamma)\right)^2$ under the constraint $\widehat{\gamma}^{\{\omega\}}_j=0$ for any $j\in\omega$ where $X_i\in\mathbb{R}^5$ is the vector composed of the first two  principal component obtained on the PCA of the environmental variables and the first three principal component obtained on the PCA of the medical variables).

\begin{table}[ht!]
\centering
\begin{tabular}{ccccc}
  \hline
Risk factor & \multicolumn{4}{c}{Estimators}\\
\cline{2-5} & $\widehat\gamma^{\{2,5\}}$ & $\widehat{\gamma}^{\{2,4,5\}}$& $\widehat{\gamma}^{\{2,3,5\}}$& $\widehat{\gamma}^{\{1,2,5\}}$\\ 
  \hline
Environmental score 1 & 0.533 & 0.640 & 0.856 & 0.000 \\ 
Environmental score 2  & 0.000 & 0.000 & 0.000 & 0.000 \\
Medical score 1 & 0.658 & 0.769 & 0.000 & 0.693 \\
Medical score 2& 0.531 & 0.000 & 0.517 & 0.721 \\ 
Medical score 3  & 0.000 & 0.000 & 0.000 & 0.000 \\ \hline
p-value & 0.322 & 0.000 & 0.000 & 0.000 \\ 
   \hline
\end{tabular}
\caption{Coefficients of the single-index regression and associated p-value obtained by the empirical likelihood ratio test.} \label{tab:EL}
\end{table}

\begin{figure}[ht!]
\centering
\includegraphics[scale=0.4]{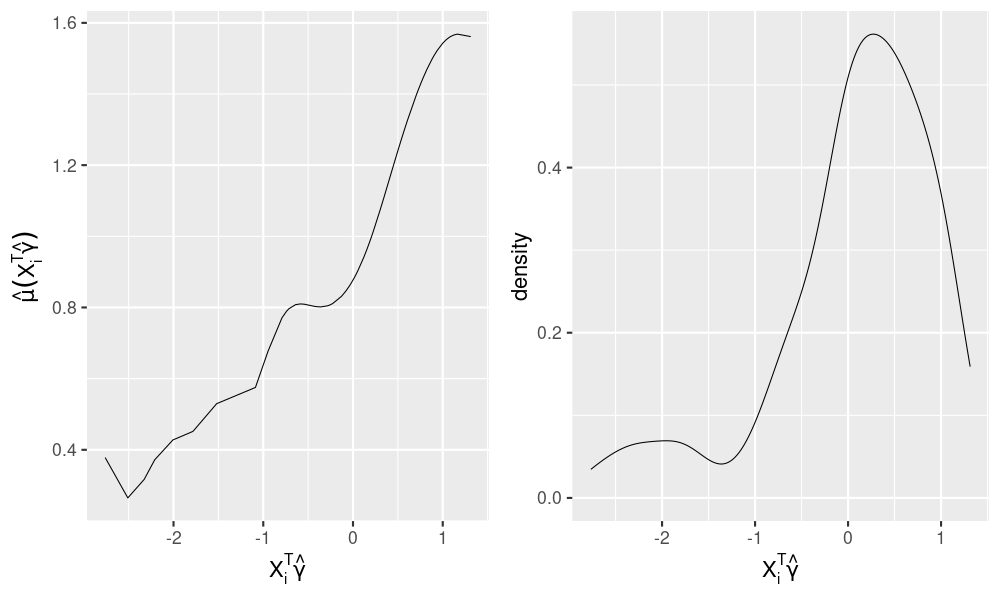}
\caption{Function $\hat\mu$ modeling the effect of the index on the \cov death curves (on the left) and density of the index $X_i^\top\hat\gamma^{\{2,5\}}$ (on the right).}
\label{fig:estimm}
\end{figure}

The impact of these covariates on the daily death rate is made throughout the index $X_i^\top\hat\gamma^{\{2,5\}}$ where $\hat\gamma^{\{2,5\}}=(0.533, 0.000, 0.658, 0.531, 0.000)^\top$  and the regression function $\hat\mu$. Figure~\ref{fig:estimm} shows the estimator of the $\hat\mu$ and the density of the index for a range covering $90\%$ of the observed index (this trimming,  performed only for this plot, avoids over-interpretation of the curve due to extreme points). Overall, the shape of $\hat\mu$ implies that the more the index is, the more is the \cov death rate. Moreover, the estimator of the density of the index is smoothed enough to ensure a consistent estimation of the regression function.  Note that the three relevant covariates in $X_i$ are composed of one PCA-scores measuring environmental risk factors and two PCA-scores measuring the medical risk factors. Thus, these results confirm that diabetes, overweight, smoking, pulmonary disease, cardiovascular disease, HIV, hypertension and age are risk factors for the \cov (see Table~\ref{tab:cor} to interpret the relation between the PCA-scores and the risk factors). These results are in agreement with the main mortality risk factors identified in medical publications (see \citet[Table 1]{zhou2020clinical} and \citet[Figure 2]{gupta2020factors}). Moreover, the population density and the concentrations of nitrogen dioxide and fine particulate matter also increase the \cov mortality risk. Additionally, these results align with the findings of other works \citep{sy2021population,copat2020role,pozzer2020regional}.

\begin{table}[ht]
\centering
\begin{tabular}{ccccccc}
  \hline
 & \multicolumn{2}{c}{Environmental score 1} &  \multicolumn{2}{c}{Medical score 1} &  \multicolumn{2}{c}{Medical score 2} \\ 
 & correlation & p-value & correlation & p-value & correlation & p-value\\ 
  \hline
PM2.5\_PopWtd & -0.566 & 0.021 & -0.435 & 0.000 & 0.199 & 0.054 \\ 
  NO2\_PopWtd & 0.796 & 0.350 & 0.307 & 0.003 & 0.041 & 0.692 \\ 
  WorldPop\_Density & 0.509 & 0.000 & -0.006 & 0.953 & -0.323 & 0.001 \\ 
  Diabetes & 0.282 & 0.423 & 0.765 & 0.000 & -0.450 & 0.000 \\ 
  Obesity & 0.334 & 0.047 & 0.853 & 0.000 & -0.153 & 0.140 \\ 
  Smoking & -0.287 & 0.131 & -0.083 & 0.426 & 0.808 & 0.000 \\ 
  COPD & 0.425 & 0.008 & 0.731 & 0.000 & 0.198 & 0.056 \\ 
  CVD & 0.301 & 0.590 & 0.846 & 0.000 & 0.068 & 0.516 \\ 
  HIV & 0.371 & 0.000 & 0.272 & 0.008 & -0.690 & 0.000 \\ 
  Hypertension & 0.196 & 0.790 & 0.821 & 0.000 & 0.344 & 0.001 \\ 
  WorldPop\_65 & 0.053 & 0.044 & 0.290 & 0.005 & 0.844 & 0.000 \\ 
   \hline
\end{tabular}
\caption{Correlation coefficients and their associated p-values for testing their nullity between the risk factors and the significant PCA scores.\label{tab:cor}}
\end{table}

\subsection{Clustering of the regions}
\label{appli:cluster}
Clustering is performed by considering a number of clusters between one and ten. Selecting the number of clusters is still a difficult task in nonparametric mixtures. Indeed, despite recent works \citep{kasahara2014non,kwon2019estimation} presenting elegant methods  for selecting the number of clusters based on the constraint of linear independence between the univariate densities required for model identifiability \citep{allman2009identifiability}, these methods require large samples to perform well (see Section~5 of \citet{kwon2019estimation}). Alternatively, \citet{DuRoy2021clustering} proposed a model selection approach that considers a discretization of each variable $\hat{\xi}_{ij}$ into bins B and a BIC penalization of the resulting log-likelihood. To ensure a consistent estimation of the number of components, the number of bins needs to tends to infinity at an appropriate rate. As suggested by the authors, we consider that $B=[n^{1/6}]$ and the bounds of the bins are defined by the empirical quantiles of  $\hat{\xi}_{ij}$. The approach permits to select $K=4$ clusters thus corresponding in the first an elbow in the values of the smoothed log-likelihood that are presented in Table~\ref{tab:loglike}. Note that a second elbow occurs and suggests that a partition into seven clusters is also of interest. In the following, we provide the interpretation of the clustering output into 4 clusters.
\begin{table}[ht]
\caption{Maximum smoothed log-likelihood $\ell(\hat\lambda_L;L)$ with respect to the number of clusters.}\label{tab:loglike}
\centering
\begin{tabular}{ccccccccccc}
  \hline
$L$ & 1 & 2 & 3 & 4 & 5 & 6 & 7 & 8 & 9 &10\\ 
$\ell(\hat\lambda_L;L)$ &--1378 & -1238 & -1188 & -1136 & -1115 & -1100 & -1066 & -1054 & -1056 & -1034\\ 
   \hline
\end{tabular}
\end{table}

\begin{figure}[ht]
\begin{center}
\includegraphics[scale=0.4]{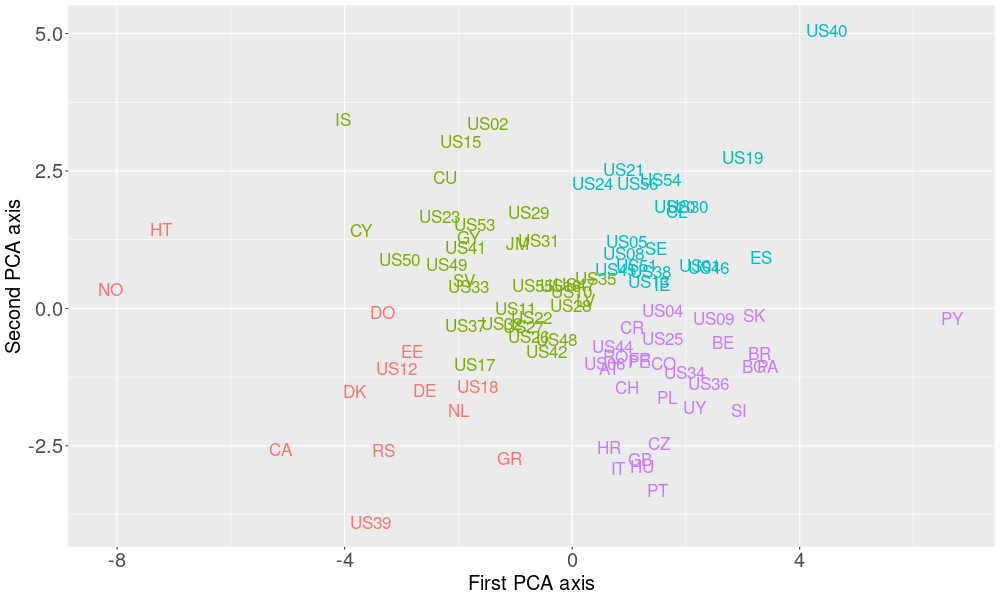}
\caption{First PCA map computed from the $\xi_1,\ldots,\xi_n$. Colors indicate the estimated partition in five classes (red: class1; maroon: class2; green: class3; blue: class4; purple: class5.} 
\label{fig:ACPmap1}
\end{center}
\end{figure}

Figure~\ref{fig:ACPmap1} represents the regions in the first PCA map computed from the $\xi_1,\ldots,\xi_n$ and it indicates the partition with colors (Table~\ref{tab:labels} presented in Appendix~\ref{app:tablabels} gives the relation between the ID of the regions and their names). Hence, it shows that the estimated partition provides no overlapping cluster in the space of the translation-invariant wavelet frame adjusted with the population risk factors. Moreover, to investigate the within-class homogeneity between the curves adjusted by the population risk factors, Table~\ref{tab:dwtdist} presents, for each class $k$ and $\ell$, the mean of DTW distances between each curve belonging to class $k$ and each curve belonging to class $\ell$ (\emph{i.e.,}  $\sum_{\{i: z_{ik}=1\}}\sum_{\{i': z_{i'\ell}=1\}} DTW(W_i(t)/\widehat{\nu}(X_i^\top\widehat{\gamma}),W_{i'}(t)/\widehat{\nu}(X_{i'}^\top\widehat{\gamma}))/n_kn_\ell$, where $n_k=\sum_{i=1}^n z_{ik}$).

\begin{table}[ht!]
\begin{center}
\begin{tabular}{ccccc}
  \hline
& Class 1& Class 4& Class 3& Class 4\\ \hline
Class 1 & 339 & 387 & 586 & 1350 \\ 
Class  2 & 387 & 371 & 535 & 1186 \\ 
Class  3 & 586 & 535 & 550 & 959 \\ 
 Class 4 & 1350 & 1186 & 959 & 1085 \\  \hline
\end{tabular}
\caption{Mean of the DTW distances between the adjusted curves of two classes.}
 \label{tab:dwtdist}
\end{center}
\end{table}

\begin{itemize}
\item Canada, Germany, Denmark, Dominican Republic, Estonia, Greece, Haiti, Netherlands, Norway, Serbia, Florida, Indiana and Ohio.
\item Cuba, Cyprus, Guyana, Iceland, Jamaica, Latvia, El Salvador, Alaska,  Delaware,  District of Columbia,  Hawaii,  Idaho,  Illinois,  Louisiana,  Maine,  Michigan,  Minnesota,  Mississippi,  Missouri,  Nebraska,  Nevada,  New Hampshire,  New Mexico,  North Carolina,  Oregon,  Pennsylvania,  Tennessee,  Texas,  Utah,  Vermont,  Washington and  Wisconsin.
\item Chile, Spain, Ireland, Sweden, Alabama, Arkansas, Colorado, Georgia, Iowa, Kansas, Kentucky, Maryland, Montana, North Dakota, Oklahoma, South Carolina, South Dakota, Virginia, West Virginia and Wyoming.
\item Austria, Belgium, Bulgaria, Brazil, Switzerland, Colombia, Costa Rica, Czechia, France, United Kingdom, Croatia, Hungary, Italy, Panama, Peru, Poland, Portugal, Paraguay, Romania, Slovenia, Slovakia, Arizona,  California,  Connecticut,  Massachusetts,  New Jersey,  New York,  Rhode Island and  Uruguay.
\end{itemize}

\begin{figure}[htp]
\begin{center}
\includegraphics[scale=0.44]{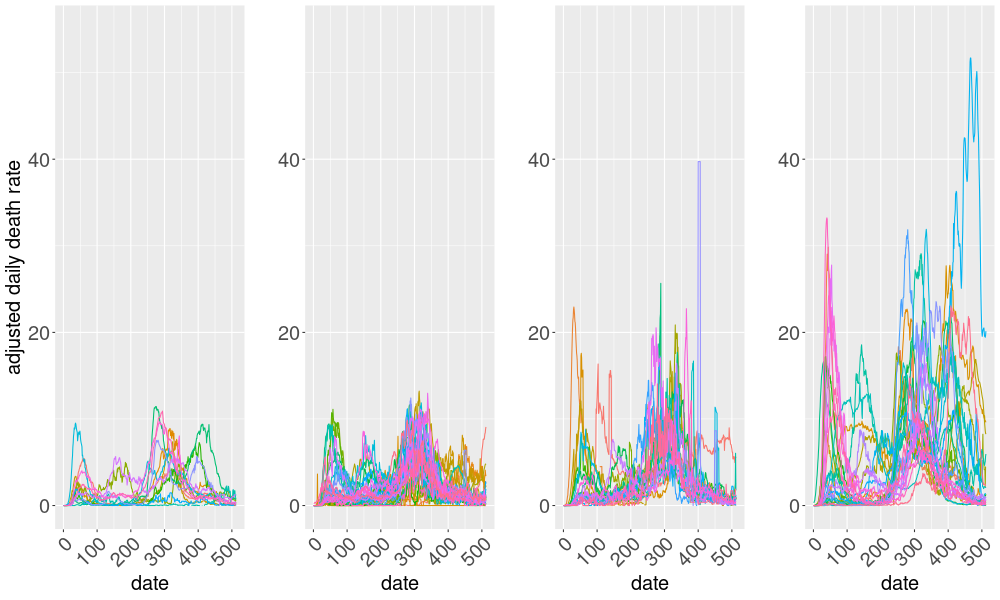}
\caption{Curves $W_{i}/\hat{\mu}(X_i^\top\hat\gamma)$ displayed per cluster (1 to 4 from the left to the right).} 
\label{fig:mapcluster}
\end{center}
\end{figure}

\begin{figure}[htp]
\begin{center}
\includegraphics[scale=0.44]{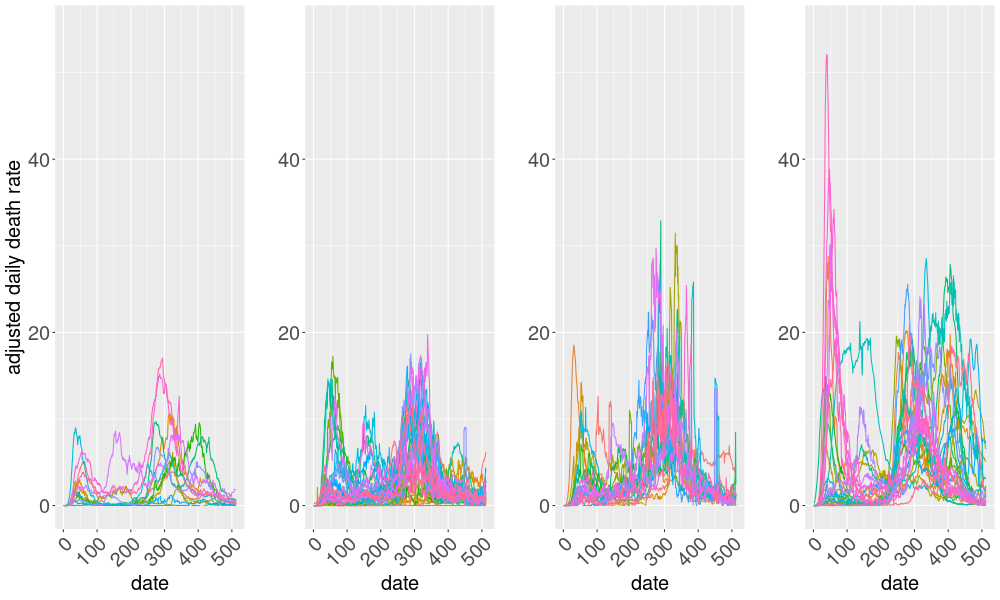}
\caption{Original curves $W_{i}$ displayed per cluster (1 to 4 from the left to the right).} 
\label{fig:mapclusterunnormalized}
\end{center}
\end{figure}

We now describe the clusters based on summary statistics presented in Table~\ref{tab:rescluster} and the curves adjusted with the population risk factors ($W_i/\hat\mu(X_i^\top\hat\gamma)$). Figure~\ref{fig:mapcluster}  presents the curves normalized by the population risk effects $W_i/\hat\mu(X_i^\top\hat\gamma)$ per cluster to illustrate our cluster interpretation. In addition, Figure~\ref{fig:mapclusterunnormalized} presents the original curves per clusters to illustrate the impact of the normalization by the population risk effects. The labels of clusters are ordered by the impact of the \cov during the studied period. Thus, we notice that cluster 1 has the smallest mean of \cov daily death rates over the studied period and cluster 5 has the highest mean. 

\begin{table}[htp]
\centering
\caption{Statistics per cluster} 
\label{tab:rescluster} 
\begin{tabular}{cccccccc}
  \hline
  cluster & proportion & \multicolumn{2}{c}{Normalized deaths} &\multicolumn{2}{c}{UnNormalized deaths} & \multicolumn{2}{c}{Covariate effect} \\
 & $\pi_\ell$ & mean & sd. & mean & sd. & mean & sd. \\ 
  \hline
Class 1 & 0.14 & 860.10 & 397.02 & 958.87 & 638.69 & 1.01 & 0.41 \\ 
Class   2 & 0.33 & 1030.55 & 414.88 & 1219.62 & 711.38 & 1.11 & 0.36 \\ 
Class   3 & 0.22 & 1463.05 & 364.75 & 1736.15 & 350.80 & 1.23 & 0.28 \\ 
Class   4 & 0.31 & 2532.97 & 961.54 & 2253.94 & 888.82 & 0.94 & 0.32 \\ 
    \hline
\end{tabular}
\end{table}

These results were expected because the distribution of $\hat\mu(X_i^\top\hat\gamma)$ is supposed to be the same among clusters. Clusters can be interpreted, as follows:
\begin{itemize}
\item Cluster 1 contains 13 regions that are mainly unaffected from the disease.
\item Cluster 2 contains 32 regions  that suffer from multiple small waves of deaths.
\item Cluster 3 contains 20 regions  mainly affected by one major wave with steep increasing and decreasing rates.
\item Cluster 4 contains 29 regions strongly impacted by the disease and that suffered for two waves at the minimum. In addition, for the regions that suffered for at least three waves of death, there is a wave of deaths which starts before the previous wave ends.
\end{itemize}

\subsection{Clusters analysis example: disparities and policy decisions}
\label{appli:supp}
Amid unforeseen difficulties related to \cov, policymakers have resorted to various interventions in attempts to curb the spread of the corona virus. As seen in Sections \ref{appli:risk} and \ref{appli:cluster}, by adjusting the risk factors, the proposed approach allows us to detect vulnerable regions and compare them fairly. Additionally, the homogeneity within cluster enables us to analyze the disparities in factors between clusters; for instance, in this paper, we are looking at the government responses. Note that one may also be interested to study \cov's impact on the population mental health or the economic indicators. 

\begin{figure}[!htp]
\centering
\subfigure[]{
\includegraphics[width=0.47\textwidth]{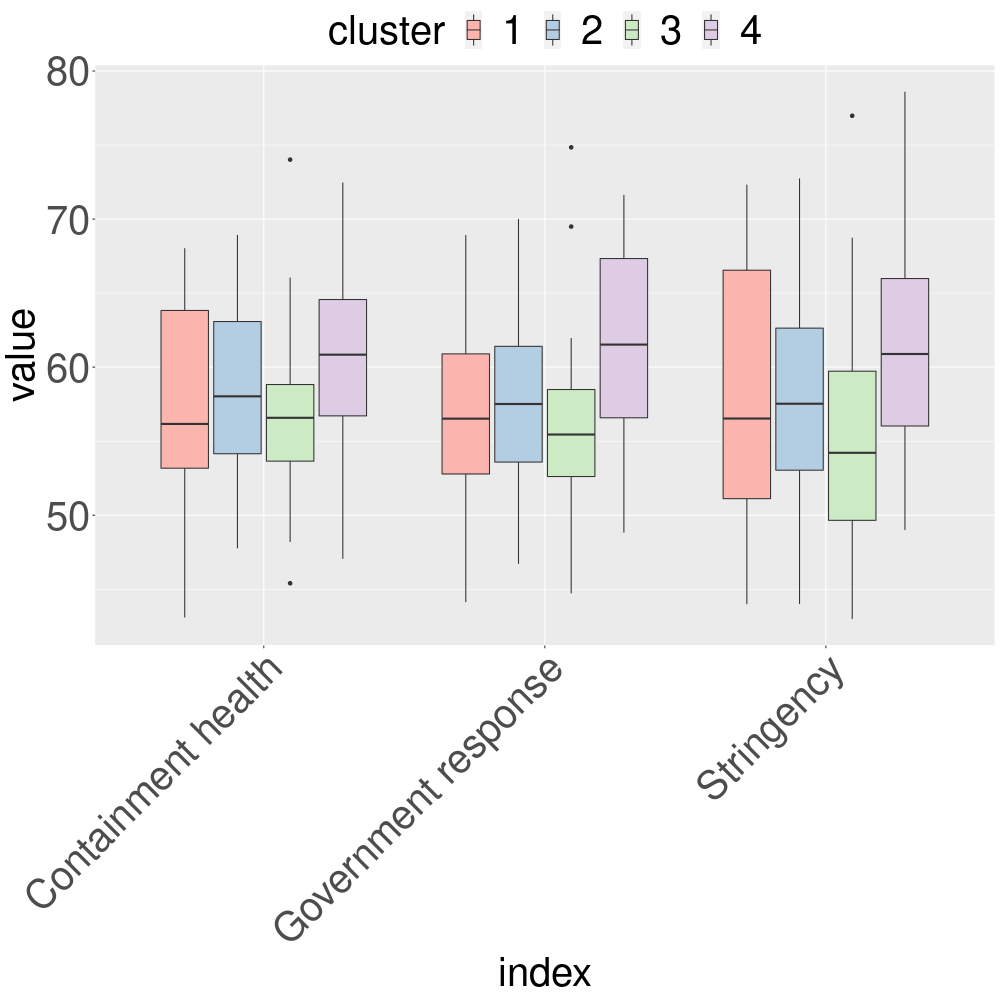}
}
\subfigure[]{
\includegraphics[width=0.47\textwidth]{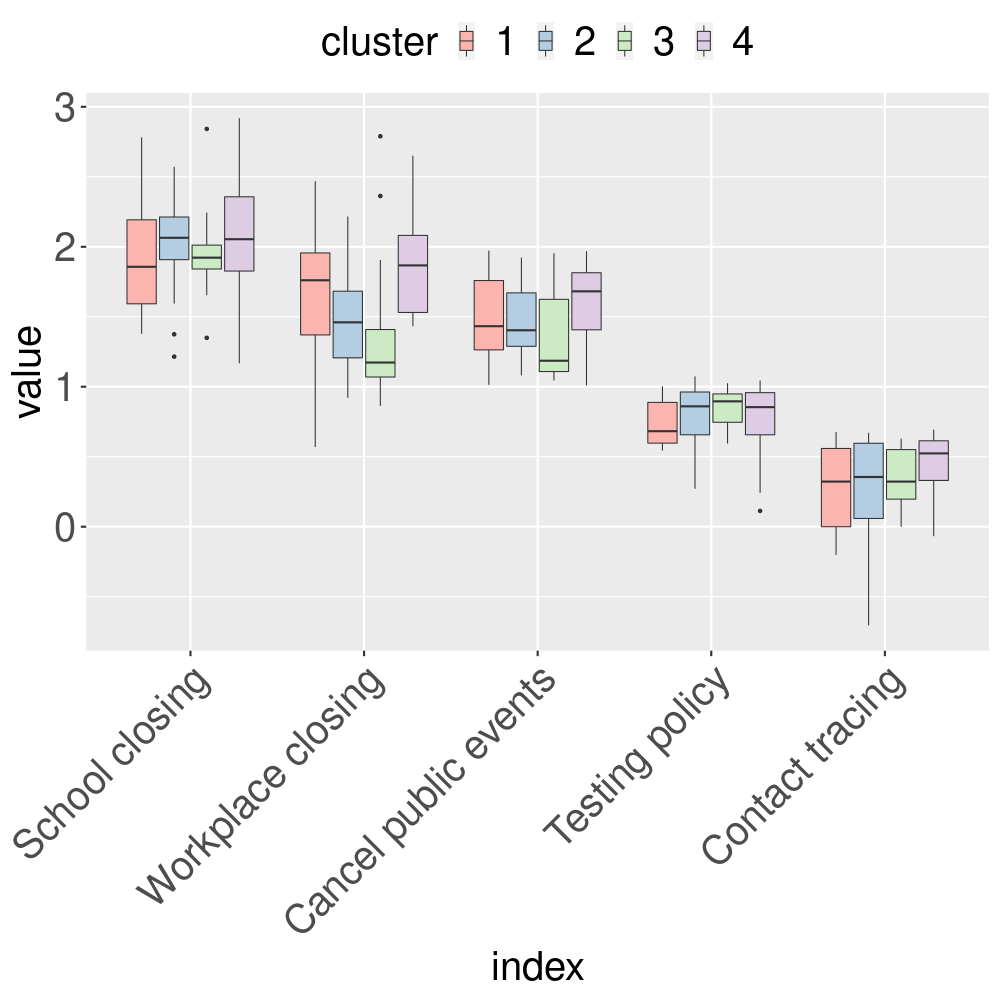}
}
\caption{Boxplot of the overall indicator values for each cluster (a) and for specific measures adopted for each cluster (b): School closing (C1), Workplace closing (C2), Cancel public events (C3), Testing policy (H2) and Contact tracing (H3).}
\label{fig:indexgrl}
\end{figure}

Figure~\ref{fig:indexgrl}(a) shows the distribution of the value for each indicator, \emph{i.e.}, containment health, government response and stringency, with respect to its clusters. We can observe that the indicators take larger values for regions of assigned in cluster 4 (containing the most impacted regions). The same phenomenon is observed for the specific measures adopted by the governments presented in Figure~\ref{fig:indexgrl}(b).

Cluster 4, which contains the regions strongly affected by the virus, seem  to take more stringent measures than the other regions. One plausible explanation is that the policymakers may abruptly enforce restrictive countermeasures in hopes to lower the increasing mortality rate, thus the observed rapid descent. Hence, the effectiveness of government response may depend on the timing of the measure's implementation, the duration and the stringency \citep{cheng2020,haug2020}. To further understand the relationship between the mortality rate and the various policies adopted in reaction to the \cov, a more precise analysis can be achieved by fitting a model for multivariate non-stationary time series per clusters (the time series being the daily death rate and the index mentioned above). Thus, a description of the relation between the government responses and the \cov death curve can be done using the model parameters. This can be achieved by modelling the dependencies between the time series \citep{molenaar1992dynamic,sanderson2010estimating} or by a multiple change-point detection \citep{cho2015multiple}. Moreover, a comparison of the government interventions to \cov can be conducted by comparing the models fitted for each cluster.

\section{Conclusion}
This paper introduces \approach, a new method to cluster functional data when observations are shifted and external covariates are allowed to have a scale effect. \approach  is a three-step approach developed for the purpose of analyzing geographical disparities of the \cov impact (measured by the daily number of deaths per million people).

As a first step of \approach, feature extraction is performed 
with \TI wavelets. 
While providing an adapted and compact representation of the data, it also permits us to deal with the different times of arrivals of the disease. The main limitation of this approach lies in the dyadic data constraint of the considered sample. 
This issue could be overcome, for example, by using second generation wavelets and in particular the lifting scheme (\cite{sweldens1998lifting}), but would lose the property of translation-invariance. However, extending this construction, while preserving both the property of translation-invariance and the property of conservation of the norm of the coefficients, seems to be an open question that we leave for future work.

As a second step of \approach, the effect of the population risk factors on the extracted feature is estimated and thus regions can be compared as if they have the same sensitivity of the population to the disease. This step is crucial because we aim to investigate the impact of policy decisions. Obviously, if the purpose is to investigate geographical disparities of the impact of the disease, then no adjustment on the population risk factors should be considered. In such case, \approach can still be used by considering $\mu(X_i^\top\gamma)=1$. In the analysis of \cov, we consider four main factors of comorbidity (overweight, diabetes, chronic pulmonary and kidney diseases). These factors are well-known to increase the risk of \cov mortality, \citep{zhou2020clinical,gupta2020factors} and Wilks' theorem for the empirical likelihood permits us to conclude on the significance of the estimated coefficients. However, we advise considering factors already known to be compounded for the disease, and not to use \approach to investigate the impact of population risk factors. Indeed, \approach does not perform variable selection of the population risk factors and does not permit concluding on causality of the factors.

As a third step of \approach, a nonparametric mixture is used to achieve the clustering with the assumption that the density of the component is defined as a product of univariate densities. Numerical experiments presented in the paper suggested that considering a more complex model deteriorated the results when the sample size is small (like in the \cov application). However, if the data to be analyzed are composed of several observations, more advanced models could be used \citep{mazo2019constraining,zhu2019clustering}.

Through the \cov dataset, we had illustrated the importance of adjusting the population risk factors, allowing us to compare regions with a `standard' comorbidity. Thus, \approach found five clusters justified by the mortality rate and curvature. Regions within clusters are varied geographically with different onsets, validating the property of translation-invariance of the proposed method. In addition, as we illustrated, investigations on  the effectiveness and agility of government response, the consequences on economic indicators or the impact on human mental health, could be achieved by studying disparities of the indicators between clusters. 

Despite the model being translation invariant, the time between the arrivals of two waves is discriminative. We argue that this time is important; it determines whether the health facilities have any breaks between waves. Indeed, countries suffering for successive waves of \cov have to postpone non-emergency surgical operation or early cancer detection. Thus, using the proposed clustering, we could investigate the impact of \cov on the global quality of care. Note that an alternative clustering approach could focus only on the death peaks thus neglecting the time between waves. In such a case, the proposed approach is not suitable and we advice to use time scaled clustering \citep{tang2009time}. 

Driven by the \cov dataset, we developed this novel approach. However, its application is not limited only to \cov. For instance, the problem of time-shifts is also observed in electrocardiogram heartbeat, which \cite{annam2011} tackles when clustering heartbeat abnormalities. Our approach could not only handle the time-shift issue, but also allow for adjusting with regards to plausible factors that may influence the heartbeat. Further, this could extend beyond medical settings: in motion capture \citep{li2011time}, there is interest in categorizing types of motion. This could be used for fitness applications to identify whether a person is running or walking, where the motions may begin at different times on separate observed sequences showing a need for the \TI property. Since motion can come from different participants, covariate adjustment could also be beneficial for such data. 


\section*{Acknowledgements}
The authors thank the editor, the associate editor, and two referees for their many constructive comments and suggestions, which greatly improved the quality of the article.	

\section*{Data Availability statement}
R code for the simulation study and data analysis, as well as dataset are included in the online supplementary materials or available from the following GitHub repository: \url{https://github.com/fabnavarro/covid-clustering}

\bibliographystyle{apalike}
\bibliography{clustfunspCovid}

\listoffigures

\appendix
\section{Proof of Lemma~1} \label{app:mixture}
\begin{proof}[Proof of Lemma~1]

We consider the matrix representation for the discrete wavelet transform. For the orthogonal case, the discrete transform of a vector $w \in \mathbb{R}^T$ is represented by an $T\times T$ orthogonal matrix $M$ (see, \emph{e.g.,} \cite{mallat:08}).
 \[
 M=
 \begin{pmatrix}
\phi(1)&\psi_{0,0}(1)&\ldots&\psi_{J-1,2^J-1}(1)\\
\vdots&\vdots& &\vdots\\
\phi(T)&\psi_{0,0}(T)&\ldots&\psi_{J-1,2^J-1}(T)\\
 \end{pmatrix}^\top.
 \]
In the translation-invariant case, the downsampling of the locations $k$ is discarded at each scale $j$, and the decomposition can thus be written as follows
\begin{equation*}
W_i(t)= \sum_{k=0}^{T-1}\alpha_{i,0,k}\phi_{0,k}(t)+\sum_{j=0}^{J-1}  \sum_{k=0}^{T-1} \beta_{i,j,k}\psi_{j,k}(t), \quad t\in [1,T], \quad i=1,\dots,n.
\end{equation*}
The matrix representation is therefore given by an $(J+1)T\times T$ matrix $M$ which corresponds to the row-wise concatenation of the $J$, $T\times T$ matrices which yield the wavelet coefficients at scale $j$ plus the matrix which produces the scaling coefficients (see, \emph{e.g.,} \cite{berkner2002smoothness,Coifman1995} for a full representation, in terms of  filters). The representation at scale $j$ for $W_i$ given $z_i$ is defined by $
M_jW_i$ where $M_j$ denotes either the $T\times 2^j$ or the $T\times T$ wavelet transform matrix at scale $j$, for $j=0,\ldots J$.
Thus, we have
\[
M_jW_i=\sum_{\ell=1}^L z_{i\ell} M_j\left[\mu(x_i)(u_\ell^{(\delta_i)} + \varepsilon_{i\ell}^{(\delta_i)}\right].
\]
Therefore, the logarithm of the norm at scale $j$ for $W_i$ given $z_i$ is given by
\[
y_{ij}=\ln\left(\mu^2(x_i)\sum_{\ell=1}^L z_{i\ell} \|v_{\ell j} + \varepsilon^\star_{i\ell j}\|_2^2\right)=\ln \mu(x_i) + \frac{1}{2}\ln\left[\sum_{\ell=1}^L\pi_{\ell} \|v_{\ell j} + \varepsilon^\star_{i\ell j}\|_2^2 \right],
\]
where $v_{\ell j}=M_ju_\ell^{(\delta_i)}$ and $\varepsilon^\star_{i\ell j}=M_j\varepsilon_{ij}^{(\delta_i)}$. 
Considering the centered version of the elements leads to \eqref{eq:SImodel}. Noting that the noise of the regression follows a mixture model where the latent variable is $z_{i}$ (\emph{i.e.}, the same latent variable that the one used to define the marginal distribution of $W_i$), permits to conclude that clustering the noise of the regression obtained for the different scale is relevant to cluster the curves $W_i$. Finally, noting that by assumption $X_i$ is independent to $Z_i$ and $\varepsilon_{i\ell}$ and that $\varepsilon_{i\ell j}^\star$ is defined as a combination of the original $\varepsilon_{i\ell}$ permits to conclude on the independence between $X_i$ and $\varepsilon_{i\ell j}^\star$.

\end{proof}

\section{Association between the region IDs and their names}
\label{app:tablabels}
\begin{table}[ht]
\centering
\begin{footnotesize}
\begin{tabular}{cccc}
  \hline
ID& Name & ID & Name \\ 
  \hline
AT & Austria & US08 & Colorado, United States \\ 
  BE & Belgium & US09 & Connecticut, United States \\ 
  BG & Bulgaria & US10 & Delaware, United States \\ 
  BR & Brazil & US11 & District of Columbia, United States \\ 
  CA & Canada & US12 & Florida, United States \\ 
  CH & Switzerland & US13 & Georgia, United States \\ 
  CL & Chile & US15 & Hawaii, United States \\ 
  CO & Colombia & US16 & Idaho, United States \\ 
  CR & Costa Rica & US17 & Illinois, United States \\ 
  CU & Cuba & US18 & Indiana, United States \\ 
  CY & Cyprus & US19 & Iowa, United States \\ 
  CZ & Czechia & US20 & Kansas, United States \\ 
  DE & Germany & US21 & Kentucky, United States \\ 
  DK & Denmark & US22 & Louisiana, United States \\ 
  DO & Dominican Republic & US23 & Maine, United States \\ 
  EE & Estonia & US24 & Maryland, United States \\ 
  ES & Spain & US25 & Massachusetts, United States \\ 
  FR & France & US26 & Michigan, United States \\ 
  GB & United Kingdom & US27 & Minnesota, United States \\ 
  GR & Greece & US28 & Mississippi, United States \\ 
  GY & Guyana & US29 & Missouri, United States \\ 
  HR & Croatia & US30 & Montana, United States \\ 
  HT & Haiti & US31 & Nebraska, United States \\ 
  HU & Hungary & US32 & Nevada, United States \\ 
  IE & Ireland & US33 & New Hampshire, United States \\ 
  IS & Iceland & US34 & New Jersey, United States \\ 
  IT & Italy & US35 & New Mexico, United States \\ 
  JM & Jamaica & US36 & New York, United States \\ 
  LV & Latvia & US37 & North Carolina, United States \\ 
  NL & Netherlands & US38 & North Dakota, United States \\ 
  NO & Norway & US39 & Ohio, United States \\ 
  PA & Panama & US40 & Oklahoma, United States \\ 
  PE & Peru & US41 & Oregon, United States \\ 
  PL & Poland & US42 & Pennsylvania, United States \\ 
  PT & Portugal & US44 & Rhode Island, United States \\ 
  PY & Paraguay & US45 & South Carolina, United States \\ 
  RO & Romania & US46 & South Dakota, United States \\ 
  RS & Serbia & US47 & Tennessee, United States \\ 
  SE & Sweden & US48 & Texas, United States \\ 
  SI & Slovenia & US49 & Utah, United States \\ 
  SK & Slovakia & US50 & Vermont, United States \\ 
  SV & El Salvador & US51 & Virginia, United States \\ 
  US01 & Alabama, United States & US53 & Washington, United States \\ 
  US02 & Alaska, United States & US54 & West Virginia, United States \\ 
  US04 & Arizona, United States & US55 & Wisconsin, United States \\ 
  US05 & Arkansas, United States & US56 & Wyoming, United States \\ 
  US06 & California, United States & UY & Uruguay \\ 
   \hline
\end{tabular}
\end{footnotesize}
\caption{ID and names of the regions. \label{tab:labels}}
\end{table}

\end{document}